\documentclass[12pt,showpacs,preprintnumbers,amsmath,amssymb]{revtex4}
\usepackage[dvips]{graphicx}

\begin{document}
\title{Flavor-changing interactions with singlet quarks
and their implications for the LHC}

\author{Katsuichi Higuchi}
\affiliation{Department of Literature, Kobe Kaisei College,
Kobe 657-0805, Japan}
\author{Katsuji Yamamoto}
\affiliation{Department of Nuclear Engineering, Kyoto University,
Kyoto 606-8501, Japan}

\date{\today}

\begin{abstract}
We investigate the flavor-changing interactions
in an extension of the standard model with singlet quarks and singlet Higgs,
which are induced by the mixing between
the ordinary quarks and the singlet quarks ($ q $-$ Q $ mixing).
We consider the effects of the gauge and scalar interactions
in the $ \Delta F = 2 $ mixings
of $ K^0 $, $ B_d $, $ B_s $ and $ D^0 $ mesons
to show the currently allowed range of the $ q $-$ Q $ mixing.
Then, we investigate the new physics
around the electroweak scale to the TeV scale,
which is accessible to the Large Hadron Collider.
Especially, the scalar coupling mediated by the singlet Higgs
may provide distinct signatures
for the decays of the singlet quarks and Higgs particles,
which should be compared with the conventionally expected ones
via the gauge and standard Higgs couplings.
Observations of the singlet quarks and Higgs particles
will present us important insights
on the $ q $-$ Q $ mixing and Higgs mixing.
\end{abstract}


\maketitle

%%%%%%%%%%
\section{Introduction}
\label{sec:intro}
%%%%%%%%%%

As the standard model has been established in current experiments,
the appearance of new physics now attracts growing interests
especially in the light of the Large Hadron Collider (LHC).
So far various extensions of the standard model
with their own motivations have been investigated for new physics,
including exotic fermions, extra Higgs fields,
extended gauge interactions, supersymmetry, and so on.
The new physics might already provide some significant effects
in the low-energy particle phenomena such as flavor-changing processes.
It is now expected seriously that the new physics will reveal itself
in the LHC experiments.

Among many intriguing extensions of the standard model,
we here investigate the new physics provided by isosinglet quarks,
which are suggested in certain models
such as $ {\rm E}_6 $-type unification
\cite{E6}.
Specifically, there are two types of singlet quarks,
$ U $ with electric charge $ Q_{\rm em} = 2/3 $
and $ D $ with $ Q_{\rm em} = - 1/3 $,
which may mix with the ordinary quarks.
It is also reasonable to incorporate a singlet Higgs field $ S $,
which provides the singlet quark masses
and the $ q $-$ Q $ mixing between the ordinary quarks ($ q = u , d $)
and the singlet quarks ($ Q = U , D $).
In this sort of model various novel features arise
through the $ q $-$ Q $ mixing
\cite{Ag,Br,BPW,R-86,ER,LL,BM,BB,qQ1,qQ2,qQ3,qQ4,qQ5,qQ6,qQ7,qQ8,FHS,
KY,HY00,HSY,qQ9,qQ10,qQ11,QB1,QB2}.
The unitarity of Cabibbo-Kobayashi-Maskawa (CKM) matrix
within the ordinary quark sector is violated,
and the flavor-changing neutral currents (FCNC's) appear
at the tree-level.
These flavor-changing interactions are described appropriately
in terms of the $ q $-$ Q $ mixing parameters and quark masses
\cite{Ag,Br,LL,FHS,HY00}.
Then, the actual CKM mixing is reproduced
up to the small unitarity violation
provided the FCNC's are suppressed sufficiently
with the small $ q $-$ Q $ mixing.
In this respect the presence of singlet quarks may introduce
an interesting extension of the notion of natural flavor conservation
\cite{GWP,AHR92HW93}.
Furthermore, the new $ CP $-violating phases in $ q $-$ Q $ mixing
may provide significant contributions
especially in the $ B $ meson physics
\cite{Ag,Br,BM,BB,qQ1,qQ3,qQ5,qQ6,KY,qQ9,qQ10,qQ11,QB1,QB2}.

It is also expected in cosmology
that the singlet quarks and singlet Higgs field
may play important roles in the early universe.
Specifically, in the first-order electroweak phase transition
the $ CP $-violating $ q $-$ Q $ mixing via the coupling
with the complex singlet Higgs $ S $ can be efficient
to produce the chiral charge fluxes through the bubble wall
for the baryogenesis \cite{ewbaryogenesis1,ewbaryogenesis2}.
Furthermore, the presence of singlet Higgs field is preferable
for realizing the strong enough first-order electroweak phase transition.

As mentioned in the above, the singlet quarks and singlet Higgs
bring various intriguing features in particle physics and cosmology.
It is hence worth considering their phenomenological implications
toward the discovery of them at the LHC
\cite{QqH-1,QqH-2,AR-04,A-S-05,SQMFV,SU-08}.
In this study we investigate the flavor-changing interactions
in the presence of singlet quarks and singlet Higgs.
The effects of the gauge interactions have been investigated
extensively in the literature
\cite{Ag,Br,BPW,R-86,ER,LL,qQ1,qQ2,qQ3,qQ4,qQ5,qQ6,qQ7,qQ8,
qQ9,qQ10,qQ11,QB1,QB2}.
Here, we rather note that the scalar interactions
mediated by the singlet Higgs may provide significant effects in some cases
\cite{BM,BB,qQ7,KY,HY00},
which has not been considered thoroughly so far
in the models with singlet quarks.

The rest of this paper is organized as follows.
In Sec. \ref {sec:mixing-FCI} we describe a representative model
with singlet quarks and one complex singlet Higgs field,
and review the essential features on the quark mixings
and flavor-changing interactions.
In Sec. \ref{sec:DF2} we consider the effects
of the flavor-changing interactions in the $ \Delta F = 2 $ mixings
of $ K^0 $, $ B_d $, $ B_s $ and $ D^0 $ mesons
to show the currently allowed range of the $ q $-$ Q $ mixing.
In Sec. \ref{sec:Qprocesses} we investigate
the decays of the singlet quarks and Higgs particles,
which may provide distinct signatures
upon their productions at the LHC.
Sec. \ref{sec:summary} is devoted to summary.
In the Appendix \ref{sec:couplings} a detailed derivation
is presented for the suitable relations
among the gauge and scalar couplings.

%%%%%%%%%%
\section{Quark mixings and flavor-changing interactions}
\label{sec:mixing-FCI}
%%%%%%%%%%

We first review the the essential features on the quark mixings
and flavor-changing interactions including the singlet quarks
and singlet Higgs, which are described appropriately
in terms of the $ q $-$ Q $ mixing parameters
and quark masses (see Ref. \cite{HY00} for the detailed description).
We consider a representative electroweak model
based on the gauge symmetry
$ {\rm SU(3)}_C \times {\rm SU(2)}_W \times {\rm U(1)}_Y $,
where singlet quarks $ U $ and $ D $
together with one complex singlet Higgs field $ S $ are incorporated.
The generic Yukawa couplings are given by
%%%%%
\begin{eqnarray}
{\cal L}_{\rm Y}
&=& - \ u^c_0 \lambda_u \Psi_{q_0} \Phi_H - U^c_0 h_u \Psi_{q_0} \Phi_H
\nonumber \\
& & - \ u^c_0 ( f_U S + f_U^\prime S^\dagger ) U_0
- U^c_0 ( \lambda_U S + \lambda_U^\prime S^\dagger ) U_0
\nonumber \\
& & - \ d^c_0 \lambda_d V_0^\dagger \Psi_{q_0} {\tilde \Phi_H}
- D^c_0 h_d V_0^\dagger \Psi_{q_0} {\tilde \Phi_H}
\nonumber \\
& & - \ d^c_0 ( f_D S + f_D^\prime S^\dagger ) D_0
- D^c_0 ( \lambda_D S + \lambda_D^\prime S^\dagger ) D_0
\nonumber \\
& & + {\rm H.c.}
\label{eqn:LYukawa}
\end{eqnarray}
%%%%%
in terms of the two-component Weyl fields
for the electroweak eigenstates with subscript ``0".
(The generation indices and Lorentz factors are omitted
here for simplicity.)
The isodoublets of left-handed ordinary quarks are represented by
%%%%%
\begin{equation}
\Psi_{q_0}
= \left( \begin{array}{c} u_0 \\ V_0 d_0 \end{array} \right)
\end{equation}
%%%%%
with a certain $ 3 \times 3 $ unitary matrix $ V_0 $.
The Higgs doublet is also given by
%%%%%
\begin{equation}
\Phi_H = \left( \begin{array}{c} H^+ \\
H^0 \end{array} \right)
\end{equation}
%%%%%
with $ {\tilde \Phi}_H \equiv i \tau_2 \Phi_H^* $.
The Higgs fields develop vacuum expectation values (VEV's),
%%%%%
\begin{equation}
\langle H^0 \rangle = v / {\sqrt 2} , \
\langle S \rangle = v_S / {\sqrt 2} ,
\end{equation}
%%%%%
where $ v = ( {\sqrt 2} G_F )^{-1/2} = 246 {\rm GeV} $,
and $ v_S \sim 100 {\rm GeV} - 1 {\rm TeV} $ is assumed.
The possible complex phase $ \delta_S $ of $ \langle S \rangle $
is not presented explicitly for simplicity of notation,
which may be effectively included
in the Yukawa couplings and the Higgs potential terms at the tree level.

%%%%%
\subsection{Quark masses and mixings}
\label{subsec:mass-mixing}
%%%%%

The quark mass matrix is produced as
%%%%%
\begin{equation}
{\cal M}_{\cal Q} = \left( \begin{array}{cc}
M_q & \Delta_{qQ} \\
   \Delta_{qQ}^\prime & M_Q \end{array} \right) .
\label{eqn:total-MQ}
\end{equation}
%%%%%
The submatrices are given by
%%%%%
\begin{eqnarray}
M_q &=& \lambda_q v / {\sqrt 2} ,
\Delta_{qQ}^\prime = h_q v / {\sqrt 2} ,
\label{eqn:subs-MQ1}
\\
\Delta_{qQ} &=& f_Q^+ v_S / {\sqrt 2} ,
M_Q = \lambda_Q^+ v_S / {\sqrt 2} ,
\label{eqn:subs-MQ2}
\end{eqnarray}
%%%%%
where
%%%%%
\begin{eqnarray}
f_Q^+ & \equiv & f_Q + f^\prime_Q ,
f_Q^- \equiv i ( f_Q - f^\prime_Q ) ,
\\
\lambda_Q^+ & \equiv & \lambda_Q + \lambda^\prime_Q ,
\lambda_Q^- \equiv i ( \lambda_Q - \lambda^\prime_Q ) .
\end{eqnarray}
%%%%%
Henceforth $ {\cal Q} = ( q , Q ) $ collectively,
and $ N_Q $ denotes the number  of singlet quarks.
The quark mass matrix $ {\cal M}_{\cal Q} $ is diagonalized
by unitary transformations $ {\cal V}_{{\cal Q}_{\rm L}} $
and $ {\cal V}_{{\cal Q}_{\rm R}} $ as
%%%%%
\begin{equation}
{\cal V}_{{\cal Q}_{\rm R}}^\dagger
{\cal M}_{\cal Q} {\cal V}_{{\cal Q}_{\rm L}}
= {\bar{\cal M}}_{\cal Q}
= \left( \begin{array}{cc}
{\bar M}_q & {\bf 0} \\ {\bf 0} & {\bar M}_Q
\end{array} \right) ,
\label{eqn:MQdiagonal}
\end{equation}
%%%%%
where
$ {\bar M}_q = {\rm diag}( m_{q_1} , m_{q_2} , m_{q_3} ) $,
$ {\bar M}_Q = {\rm diag}( m_{Q_1} , \ldots ) $,
and $ ( q_1 , q_2 , q_3 ) = ( u , c , t ) $ or $ ( d , s , b ) $.
The quark mass eigenstates $ q_i $ ($ i = 1, 2, 3 $)
and $ Q_a $ ($ a = 1, 2, \ldots , N_Q $) are given by
%%%%%
\begin{eqnarray}
\left( \begin{array}{c} q \\ Q \end{array} \right)
= {\cal V}_{{\cal Q}_{\rm L}}^\dagger
\left( \begin{array}{c} q_0 \\ Q_0 \end{array} \right) ,
( q^c , Q^c ) = ( q_0^c , Q_0^c ){\cal V}_{{\cal Q}_{\rm R}}
\end{eqnarray}
%%%%%
with the $ ( 3 + N_Q ) \times ( 3 + N_Q ) $ unitary matrices as
%%%%%
\begin{equation}
{\cal V}_{{\cal Q}_\chi}
= \left( \begin{array}{cc}
V_{q_\chi} & \epsilon_{q_\chi} \\
- \epsilon_{q_\chi}^{\prime \dagger} & V_{Q_\chi}
\end{array} \right) ( \chi = {\rm L} , {\rm R} ) ,
\label{eqn:VQchi}
\end{equation}
%%%%%
where $ \epsilon_{q_\chi} $ and $ \epsilon_{q_\chi}^\prime $
represent the $ q $-$ Q $ mixing.

The quark mass matrix $ {\cal M}_{\cal Q} $
may be reduced to a specific form
with either $ \Delta_{qQ}^\prime = {\bf 0} $
or $ \Delta_{qQ} = {\bf 0} $ by a unitary transformation
of the right-handed quarks.
Then, the Yukawa coupling $ \lambda_q $ is diagonalized
by unitary transformations of the ordinary quarks as
%%%%%
\begin{equation}
\lambda_q = {\rm diag}( \lambda_{q_1} , \lambda_{q_2} , \lambda_{q_3} ) ,
\label{eqn:lmq}
\end{equation}
%%%%%
while the condition $ \Delta_{qQ}^\prime = {\bf 0} $
or $ \Delta_{qQ} = {\bf 0} $ is maintained.
These transformations to specify the form of $ {\cal M}_{\cal Q} $
do not mix the electroweak doublets with the singlets,
respecting the $ {\rm SU(3)}_C \times {\rm SU(2)}_W \times {\rm U(1)}_Y $.
Hence, without loss of generality we may start with either of these bases,
%%%%%
\begin{eqnarray}
{\mbox{\bf basis (a)}} : \Delta^\prime_{qQ} = {\bf 0} ,
{\mbox{\bf basis (b)}} : \Delta_{qQ} = {\bf 0} .
\nonumber
\end{eqnarray}
%%%%%
The Yukawa couplings $ h_q $, $ f_Q $, $ f_Q^\prime $,
$ \lambda_Q $, $ \lambda_Q^\prime $
and the mixing matrix $ V_0 $ are redefined
according to the transformations to specify the quark basis.
In particular, $ h_q = {\bf 0} $ solely in the basis (a).
On the other hand, in the basis (b)
a specific relation $ f_Q^\prime = - f_Q $ ($ f_Q^+ = {\bf 0} $)
holds apparently though no tuning is imposed
among the couplings in the original basis.

The $ q $-$ Q $ mixings are given specifically in the basis (a) as
%%%%%
\begin{eqnarray}
( \epsilon_{q_{\rm L}} )_{ia} & \sim & ( \epsilon_{q_{\rm L}}^\prime )_{ia}
\sim ( m_{q_i} / m_Q ) \epsilon^f_i ,
\label{eqn:eqL-a} \\
( \epsilon_{q_{\rm R}} )_{ia} & \sim & ( \epsilon_{q_{\rm R}}^\prime )_{ia}
\sim \epsilon^f_i
\label{eqn:epR-a}
\end{eqnarray}
%%%%%
in terms of the $ q $-$ Q $ mixing parameters from the $ f_Q^+ $ coupling,
%%%%%
\begin{eqnarray}
\epsilon^f_i = ( v_S / m_Q ) {\overline{| ( f_Q^+ )_{ia} |}} / {\sqrt 2}
= {\overline{| ( \Delta_{qQ} )_{ia} |}} / m_Q ,
\label{eqn:epf}
\end{eqnarray}
%%%%%
where $ m_Q = {\overline{m_{Q_a}}}
\sim ( {\overline{| ( f_Q^+ )_{ia} |}}
+ {\overline{| ( \lambda_Q^+ )_{ab} |}} ) v_S $,
and the bar denotes the mean value.
The left-handed $ q $-$ Q $ mixing is suppressed significantly
by the $ q/Q $ mass ratios $ m_{q_i} / m_Q $
\cite{Ag,Br,LL,FHS,HY00}.
On the other hand, in the basis (b)
%%%%%
\begin{eqnarray}
( \epsilon_{q_{\rm L}} )_{ia} & \sim & ( \epsilon_{q_{\rm L}}^\prime )_{ia}
\sim \epsilon^h_i ,
\label{eqn:eqL-b} \\
( \epsilon_{q_{\rm R}} )_{ia} & \sim & ( \epsilon_{q_{\rm R}}^\prime )_{ia}
\sim ( m_{q_i} / m_Q ) \epsilon^h_i
\label{eqn:eqR-b}
\end{eqnarray}
%%%%%
in terms of the $ q $-$ Q $ mixing parameters from the $ h_q $ coupling,
%%%%%
\begin{eqnarray}
\epsilon^h_i = ( v / m_Q ) {\overline{| ( h_q )_{ai} |}} / {\sqrt 2}
= {\overline{| ( \Delta_{qQ}^\prime )_{ai} |}} / m_Q .
\label{eqn:eph}
\end{eqnarray}
%%%%%
The left-handed $ q $-$ Q $ mixing is no longer suppressed
by the $ q/Q $ mass ratios.

We may move from the basis (a) with $ \Delta_{qQ}^\prime = {\bf 0} $
to the basis (b) with $ \Delta_{qQ} = {\bf 0} $
by using a unitary transformation.
Here, the left-handed $ q $-$ Q $ mixings in the bases (a) and (b)
are related as $ \epsilon^h_i \sim ( m_{q_i} / m_Q ) \epsilon^f_i $
so that Eq. (\ref{eqn:eqL-a}) is apparently reproduced
from Eq. (\ref{eqn:eqL-b}).
Hence, the basis (a) may be regarded as a special case of the basis (b)
\cite{HY00}.
If $ {\overline{| f_Q |}} + {\overline{| f_Q^\prime |}}
\gtrsim {\overline{| \lambda_Q |}} + {\overline{| \lambda_Q^\prime |}} $
providing $ \epsilon^f_i \sim 1 $ in the basis (a),
then it is suitable to adopt the basis (b) alternatively.
The see-saw basis with $ M_q = {\bf 0} $ is also possible for $ N_Q = 3 $
\cite{qseesaw}.
Since it is related to the bases (a) and (b)
having a hybrid feature for the quark mixing \cite{HY00},
we do not consider explicitly the see-saw model.
We adopt complementarily the bases (a) and (b),
where the ordinary quark masses are reproduced as
%%%%%
\begin{equation}
m_{q_i} = c_i \lambda_{q_i} v / {\sqrt 2}
\end{equation}
%%%%%
with $ c_i  \sim 1 $ depending on the small $ q $-$ Q $ mixing
\cite{Ag,Br,LL,FHS,HY00}.
In the general basis with $ \Delta_{qQ} \not= {\bf 0} $
and $ \Delta_{qQ}^\prime \not= {\bf 0} $, e.g., the see-saw model,
the ordinary quark mass hierarchy is not described clearly
in terms of the Yukawa couplings $ \lambda_{q_i} $.

%%%%%
\subsection{Flavor-changing interactions}
\label{subsec:FCI}
%%%%%

The CKM matrix $ V $ for the $ W $-boson coupling with the ordinary quarks
is given by
%%%%%
\begin{equation}
V = V_{u_{\rm L}}^\dagger V_0 V_{d_{\rm L}} ,
\label{eqn:VCKM}
\end{equation}
%%%%%
where $ V_{u_{\rm L}} $ and $ V_{d_{\rm L}} $
are the $ 3 \times 3 $ submatrices in Eq. (\ref{eqn:VQchi}).
The unitarity violation of $ V $ is induced
at the second order of $ q $-$ Q $ mixing
with $ \epsilon_{q_{\rm L}} \epsilon_{q_{\rm L}}^\dagger $ and
$ \epsilon_{q_{\rm L}}^\prime \epsilon_{q_{\rm L}}^{\prime \dagger} $,
which should be suppressed enough phenomenologically
\cite{Ag,Br,BPW,R-86,ER,LL,BM,BB,qQ1,qQ2,qQ3,qQ4,qQ5,qQ6,qQ7,qQ8,FHS,
KY,HY00,HSY,qQ9,qQ10,qQ11,QB1,QB2}.
Then, the realistic CKM matrix $ V $ is reproduced
by taking suitably the original $ V_0 $.

The modification of the left-handed $ Z $-boson coupling
with the ordinary quarks is also given
at the second order of $ q $-$ Q $ mixing
\cite{Ag,Br,BPW,R-86,ER,LL,BM,BB,qQ1,qQ2,qQ3,qQ4,qQ5,qQ6,qQ7,qQ8,FHS,
KY,HY00,HSY,qQ9,qQ10,qQ11,QB1,QB2}
as
%%%%%
\begin{equation}
\Delta {\cal Z}_{\cal Q} [q^\dagger q]
= - \epsilon_{q_{\rm L}}^\prime
\epsilon_{q_{\rm L}}^{\prime \dagger} ,
\label{eqn:DZQ}
\end{equation}
%%%%%
where the $ Z $-boson coupling is presented
by removing the isospin factor $ I_3 (q_0) $
with $ I_3 (u_0) = 1/2 $ and $ I_3 (d_0) = -1/2 $
for simplicity of notation.
The right-handed coupling is, on the other hand, unchanged
as $ \Delta {\cal Z}_{{\cal Q}^c} = {\bf 0} $
for $ I_3 ( q^c_0 ) = I_3 ( Q^c_0 ) = 0 $.
Specifically, in the basis (a) we have
%%%%%
\begin{equation}
\Delta {\cal Z}_{\cal Q} [q^\dagger q]_{ij} {\rm (a)}
\sim ( m_{q_i} / m_Q ) ( m_{q_j} / m_Q ) \epsilon^f_i \epsilon^f_j .
\label{eqn:DZQ-a}
\end{equation}
%%%%%
This correction as well as the CKM unitarity violation
are suppressed substantially by the second order of $ q/Q $ mass ratios.
Alternatively, in the basis (b) we have
%%%%%
\begin{equation}
\Delta {\cal Z}_{\cal Q} [q^\dagger q]_{ij} {\rm (b)}
\sim \epsilon^h_i \epsilon^h_j ,
\label{eqn:DZQ-b}
\end{equation}
%%%%%
which is no longer suppressed by the $ q/Q $ mass ratios.
Then, significant constraints are placed phenomenologically
on the $ q $-$ Q $ mixings $ \epsilon^h_i \ll 1 $,
which have been investigated extensively in the literature
\cite{qQ1,qQ2,qQ3,qQ4,qQ5,qQ6,qQ7,qQ8,qQ9,qQ10}.

The neutral scalar couplings of the quarks $ {\cal Q} = {\cal U}, {\cal D} $
are extracted from Eq. (\ref{eqn:LYukawa}) as
%%%%%
\begin{equation}
{\cal L}_\phi ({\cal Q})
= - \sum_{\phi^0_r = H, S_+ , S_-}
{\cal Q}^c \Lambda_{\cal Q}^{\phi^0_r} {\cal Q} \phi^0_r
+ {\rm h.c.} ,
\end{equation}
%%%%%
where
%%%%%
\begin{eqnarray}
\Lambda_{\cal Q}^H
&=& \frac{1}{\sqrt 2} {\cal V}_{{\cal Q}_{\rm R}}^\dagger
\left( \begin{array}{cc} \lambda_q & {\bf 0} \\
h_q & {\bf 0} \end{array} \right)
{\cal V}_{{\cal Q}_{\rm L}} ,
\label{eqn:LmH}
\\
\Lambda_{\cal Q}^{S_\pm}
&=& \frac{1}{\sqrt 2} {\cal V}_{{\cal Q}_{\rm R}}^\dagger
\left( \begin{array}{cc}
{\bf 0} & f_Q^\pm \\
{\bf 0} & \lambda_Q^\pm
\end{array} \right) {\cal V}_{{\cal Q}_{\rm L}} .
\label{eqn:LmS}
\end{eqnarray}
%%%%%
The real neutral scalar fields,
$ \phi^0_1 = H \equiv {\sqrt 2} {\rm Re} ( H^0 - \langle H^0 \rangle ) $,
$ \phi^0_2 = S_+ \equiv {\sqrt 2} {\rm Re} ( S - \langle S \rangle ) $,
$ \phi^0_3 = S_- \equiv {\sqrt 2} {\rm Im} ( S - \langle S \rangle ) $,
mix generally to form the mass eigenstates $ \phi_r $
($ r = 1,2,3 $) through an orthogonal transformation $ O_\phi $:
%%%%%
\begin{equation}
\left( \begin{array}{c} \phi_1 \\ \phi_2 \\ \phi_3 \end{array} \right)
= O_\phi \left( \begin{array}{c} H \\ S_+ \\ S_- \end{array} \right) .
\end{equation}
%%%%%
The Nambu-Goldstone mode
$ G \equiv {\sqrt 2} {\rm Im} ( H^0 - \langle H^0 \rangle ) $
is absorbed by the $ Z $ boson.

The submatrices $ \Lambda_{\cal Q}^{\phi^0_r} [q^c q] $
of these neutral scalar couplings for the ordinary quarks are given by
%%%%%
\begin{eqnarray}
\Lambda_{\cal Q}^H [q^c q]
&=& V_{q_{\rm R}}^\dagger \lambda_q V_{q_{\rm L}}
- \epsilon_{q_{\rm R}}^\prime h_q V_{q_{\rm L}} ,
\label{eqn:LmHqq}
\\
\Lambda_{\cal Q}^{S_\pm} [q^c q]
&=& - V_{q_{\rm R}}^\dagger f_Q^\pm \epsilon_{q_{\rm L}}^{\prime \dagger}
+ \epsilon_{q_{\rm R}}^\prime \lambda_Q^\pm
\epsilon_{q_{\rm L}}^{\prime \dagger} .
\label{eqn:LmSqq}
\end{eqnarray}
%%%%%
Here, some close relations hold for the gauge and scalar couplings
(see the Appendix \ref{sec:couplings} for derivation).
The coupling of the standard Higgs $ H $ is given actually as
%%%%
\begin{eqnarray}
\Lambda_{\cal Q}^H [q^c q]_{ij}
&=& ( m_{q_i} / v )
( \delta_{ij} + \Delta {\cal Z}_{\cal Q} [q^\dagger q]_{ij} )
\label{eqn:lmH-DZQ}
\end{eqnarray}
%%%%%
with the $ q $-$ Q $  mixing induced $ Z $-boson coupling
$ \Delta {\cal Z}_{\cal Q} [q^\dagger q] $ in Eq. (\ref{eqn:DZQ}).
Similarly, the coupling of the singlet Higgs $ S_+ $ is calculated as
%%%%%
\begin{eqnarray}
\Lambda_{\cal Q}^{S_+} [q^c q]_{ij}
= - ( m_{q_i} / v_S ) \Delta {\cal Z}_{\cal Q} [q^\dagger q]_{ij} .
\label{eqn:LmSp}
\end{eqnarray}
%%%%%
Hence the $ q $-$ Q $ mixing effects for the scalar couplings
$ \Lambda_{\cal Q}^H [q^c q] $ and $ \Lambda_{\cal Q}^{S_+} [q^c q] $
are always sub-leading compared with those for the $ Z $-boson coupling
$ \Delta {\cal Z}_{\cal Q} [q^\dagger q] $,
which is due to the suppression by $ m_{q_i} / v $ from the chirality flip.
It is rather remarkable that the coupling of the singlet Higgs $ S_- $
may be dominant without a close relation to the $ Z $-boson coupling.
In the basis (a) we have
%%%%%
\begin{eqnarray}
\Lambda_{\cal Q}^{S_-} [q^c q]_{ij} {\rm (a)}
\sim ( m_{q_j} / v_S ) \epsilon^f_i \epsilon^f_j ,
\label{eqn:LmSma}
\end{eqnarray}
%%%%%
where $ ( \epsilon_{q_{\rm L}}^{\prime \dagger} )_{aj}
\sim ( m_{q_j} / m_Q ) \epsilon^f_j $,
$ ( \epsilon_{q_{\rm R}}^\prime )_{ia} \sim \epsilon^f_i $,
$ ( \lambda_Q^- )_{ab} \sim ( m_Q / v_S ) $
and $ ( f_Q^- )_{ia} \sim ( m_Q / v_S ) \epsilon^f_i $
are applied in Eq. (\ref{eqn:LmSqq}).
In contrast to the $ Z $-boson coupling
$ \Delta {\cal Z}_{\cal Q} [q^\dagger q] $ in Eq. (\ref{eqn:DZQ-a}),
the scalar coupling $ \Lambda_{\cal Q}^{S_-} [q^c q] $
in Eq. (\ref{eqn:LmSma}) is suppressed
only by the first order of ordinary quark mass.
In the basis (b), by applying
$ ( \epsilon_{q_{\rm L}}^{\prime \dagger} )_{aj} \sim \epsilon^h_j $
and $ ( f_Q^- )_{ia} \sim ( m_Q / v_S ) \epsilon^f_i $
in Eq. (\ref{eqn:LmSqq}), we estimate
%%%%%
\begin{eqnarray}
\Lambda_{\cal Q}^{S_-} [q^c q]_{ij} {\rm (b)}
& \sim & ( m_Q / v_S ) \epsilon^f_i \epsilon^h_j ,
\label{eqn:LmSmb}
\end{eqnarray}
%%%%%
up to the sub-leading contribution of the second term
$ \sim ( m_{q_i} / v_S ) \epsilon^h_i \epsilon^h_j $
in Eq. (\ref{eqn:LmSqq}).
Here, similar to Eq. (\ref{eqn:epf}), the $ q $-$ Q $ mixing parameters
are introduced for convenience as
$ \epsilon^f_i = ( v_S / m_Q ) {\overline{| 2 ( f_Q )_{ia} |}}/{\sqrt 2} $
even though $ f_Q^+ = {\bf 0} $
($ f_Q^- = 2i f_Q $ with $ f_Q = - f_Q^\prime $)
for $ \Delta_{qQ} = {\bf 0} $ in the basis (b).
It should also be noted, as discussed previously,
that by considering the relation for the left-handed $ q $-$ Q $ mixing,
%%%%%
\begin{eqnarray}
\epsilon^h_i \sim ( m_{q_i} / m_Q ) \epsilon^f_i ,
\label{eqn:(a)-(b)}
\end{eqnarray}
%%%%%
Eqs. (\ref{eqn:DZQ-b}) and (\ref{eqn:LmSmb}) in the basis (b)
reproduce Eqs. (\ref{eqn:DZQ-a}) and (\ref{eqn:LmSma}) in the basis (a),
respectively.

We mention for completeness
that in the case of one real $ S $ (or one supersymmetric $ S $)
with the $ f_Q $ and $ \lambda_Q $ couplings
($ f_Q^\prime \equiv {\bf 0} $ and $ \lambda_Q^\prime \equiv {\bf 0} $),
the scalar coupling $ \Lambda_{\cal Q}^S [q^c q] $
is given by Eq. (\ref{eqn:LmSp})
for $ \Lambda_{\cal Q}^{S_+} [q^c q] $
related to the $ Z $-boson coupling
$ \Delta {\cal Z}_{\cal Q} [q^\dagger q] $.
On the other hand, if the bare mass term $ M_Q $ is adopted
instead of the $ \lambda_Q $ coupling
while the $ f_Q $ coupling provides the $ q $-$ Q $ mixing,
the scalar coupling $ \Lambda_{\cal Q}^{S} [q^c q] $
is rather given by Eqs. (\ref{eqn:LmSma}) and (\ref{eqn:LmSmb})
for $ \Lambda_{\cal Q}^{S_-} [q^c q] $
even in the case of one real $ S $.

%%%%%%%%%%
\section{Singlet quark Effects in $ \Delta F = 2 $ mixings
of neutral mesons}
\label{sec:DF2}
%%%%%%%%%%

We perform a detailed analysis
on the $ q $-$ Q $ mixing effects
in the $ \Delta F = 2 $ mixings of $ K^0 $, $ B_d $, $ B_s $
and $ D^0 $ mesons, by considering the general bounds for new physics
which are presented in Ref. \cite{UTfit}.
The $ Z $-mediated FCNC's in $ \Delta {\cal Z}_{\cal Q} [q^\dagger q] $
have been investigated extensively in the literature
\cite{Ag,Br,BPW,R-86,ER,LL,qQ1,qQ2,qQ3,qQ4,qQ5,qQ6,qQ7,qQ8,
qQ9,qQ10,qQ11,QB1,QB2}.
By placing experimental constraints
on the left-handed $ q $-$ Q $ mixings
$ ( \epsilon_{q_{\rm L}} )_{ia} \sim ( \epsilon_{q_{\rm L}}^\prime )_{ia}
\sim \epsilon^h_i $ in the basis (b),
these analyses have discussed the possibility of new physics
provided by the singlet quarks, in particular, for the $ B $ meson physics.
Here, we rather note that in some cases
the scalar FCNC's in $ \Lambda_{\cal Q}^{S_-} [q^c q] $
may dominate over the $ Z $-mediated FCNC's
in $ \Delta {\cal Z}_{\cal Q} [q^\dagger q] $,
providing distinct signals for new physics
\cite{BM,BB,qQ7,KY,HY00}.
This intriguing possibility has not been paid so much attention so far
in the models with singlet quarks.

The effective Hamiltonian contributing
to the $ \Delta F = 2 $ mixing 
of the neutral meson $ M $ ($ K^0 , B_d , B_s , D^0 $)
is given generally \cite {UTfit} as
%%%%%
\begin{eqnarray}
{\cal H}^{\Delta F = 2}_{\rm eff}
= \sum_{k=1}^5 C_M^k {\cal O}^{q_i q_j}_k
+ \sum_{k=1}^3 {\tilde C}_M^k {\tilde{\cal O}}^{q_i q_j}_k ,
\end{eqnarray}
%%%%%
where
%%%%%
\begin{eqnarray}
q_i q_j = sd ( K^0 ) , bd ( B_d ) , bs ( B_s ) , cu ( D^0 ) .
\end{eqnarray}
%%%%%
The four-quark operators are
%%%%%
\begin{eqnarray}
{\cal O}^{q_i q_j}_1
&=& {\bar q}^\alpha_{j{\rm L}} \gamma_\mu q^\alpha_{i{\rm L}}
{\bar q}^\beta_{j{\rm L}} \gamma^\mu q^\beta_{i{\rm L}} ,
{\cal O}^{q_i q_j}_2
= {\bar q}^\alpha_{j{\rm R}} q^\alpha_{i{\rm L}}
{\bar q}^\beta_{j{\rm R}} q^\beta_{i{\rm L}} ,
\nonumber \\
{\cal O}^{q_i q_j}_3
&=& {\bar q}^\alpha_{j{\rm R}} q^\beta_{i{\rm L}}
{\bar q}^\beta_{j{\rm R}} q^\alpha_{i{\rm L}} ,
{\cal O}^{q_i q_j}_4
= {\bar q}^\alpha_{j{\rm R}} q^\alpha_{i{\rm L}}
{\bar q}^\beta_{j{\rm L}} q^\beta_{i{\rm R}} ,
\nonumber \\
{\cal O}^{q_i q_j}_5
&=& {\bar q}^\alpha_{j{\rm R}} q^\beta_{i{\rm L}}
{\bar q}^\beta_{j{\rm L}} q^\alpha_{i{\rm R}} ,
\nonumber
\end{eqnarray}
%%%%%
and $ \alpha $ and $ \beta $ denote the colors.
The operators $ {\tilde{\cal O}}^{q_i q_j}_{1,2,3} $
are obtained from the operators $ {\cal O}^{q_i q_j}_{1,2,3} $
by the exchange $ {\rm L} \leftrightarrow {\rm R} $.
The coefficients in the effective Hamiltonian
at the scale $ \mu = m_Q $ of singlet quarks are calculated as
%%%%%
\begin{eqnarray}
C_M^1 ( m_Q )
&=& ( g / 2 \cos \theta_W )^2
( \Delta {\cal Z}_{\cal Q} [q^\dagger q]_{ji} )^2 / m_Z^2 ,
\label{eqn:C1}
\\
C_M^2 ( m_Q )
&=& ( \Lambda_{\cal Q}^{S_-} [q^c q]_{ji} )^2 / m_{S_-}^2 ,
\label{eqn:C2}
\\
{\tilde C}_M^2 ( m_Q )
&=& ( \Lambda_{\cal Q}^{S_-} [q^c q]_{ij}^* )^2 / m_{S_-}^2 ,
\label{eqn:Ct2}
\\
C_M^4 ( m_Q )
&=& ( \Lambda_{\cal Q}^{S_-} [q^c q]_{ji} )
( \Lambda_{\cal Q}^{S_-} [q^c q]_{ij}^* ) / m_{S_-}^2 ,
\label{eqn:C4}
\end{eqnarray}
%%%%%
and the others are zero.
Here, the $ Z $-boson coupling $ \Delta {\cal Z}_{\cal Q} [q^\dagger q] $
and the dominant scalar coupling $ \Lambda_{\cal Q}^{S_-} [q^c q] $
are considered, and the scalar mixing in $ O_\phi $
is neglected for simplicity.
By requiring that these coefficients
in Eqs. (\ref{eqn:C1}) -- (\ref{eqn:C4}) are all within the bounds
presented specifically in Table 4 of Ref. \cite{UTfit},
we find the allowed range of the $ q $-$ Q $ mixing
depending on the masses $ m_Q , m_{S_-} \sim v_S $
of the singlet quarks $ Q $ and singlet Higgs $ S_- $.

The constraints on the $ q $-$ Q $ mixing are given roughly below,
where $ m_D = m_U = v_S = 500 {\rm GeV} $
and $ m_{S_-} = 0.6 v_S = 300 {\rm GeV} $
are taken typically to estimate the FCNC's
with Eqs. (\ref{eqn:DZQ-a}), (\ref{eqn:DZQ-b}),
(\ref{eqn:LmSma}) and (\ref{eqn:LmSmb})
in terms of the $ q $-$ Q $ mixing parameters
$ \epsilon^f_i $ and $ \epsilon^h_i $.
In the basis (a) significant constraints on the $ d $-$ D $ mixing
are placed by the scalar coupling $ \Lambda_{\cal D}^{S_-} [d^c d] $
for $ C_M^2 $, $ {\tilde C}_M^2 $ and $ C_M^4 $
($ M = K^0 , B_d , B_d $) as
%%%%%
\begin{eqnarray}
\Lambda_{\cal D}^{S_-}{\mbox{(a)}} :
\begin{array}{ll}
( \epsilon^f_1 \epsilon^f_2 )^{\frac{1}{2}}
\lesssim 0.1 / \delta_{12}^{\frac{1}{4}} & ( {\rm Im} K )
\\
( \epsilon^f_1 \epsilon^f_2 )^{\frac{1}{2}} \lesssim 0.4 & ( {\rm Re} K )
\\
( \epsilon^f_1 \epsilon^f_3 )^{\frac{1}{2}} \lesssim 0.2 & ( | B_d | )
\\
( \epsilon^f_2 \epsilon^f_3 )^{\frac{1}{2}} \lesssim 0.5 & ( | B_s | ) .
\end{array}
\label{eqn:efLmSma}
\end{eqnarray}
%%%%%
Here, the effective $ CP $-violating phases in the FCNC's
contributing to the $ K^0 $-$ {\bar K}^0 $ mixing
are denoted collectively by $ \delta_{12} $.
No significant constraints are, on the other hand,
placed by the $ Z $-boson coupling
$ \Delta {\cal Z}_{\cal D} [d^\dagger d] $
which is substantially suppressed by the second order
of $ d $/$ D $ mass ratios in Eq. (\ref{eqn:DZQ-a}).
Alternatively, in the basis (b) constraints
on the $ d $-$ D $ mixing are given as
%%%%%
\begin{eqnarray}
&{}& \Lambda_{\cal D}^{S_-}{\mbox{(b)}} :
\begin{array}{ll}
( \epsilon^f_1 \epsilon^h_2 )^{\frac{1}{2}} ,
( \epsilon^f_2 \epsilon^h_1 )^{\frac{1}{2}}
\lesssim 1 \times 10^{-3} / \delta_{12}^{\frac{1}{4}} &  ( {\rm Im} K )
\\
( \epsilon^f_1 \epsilon^h_2 )^{\frac{1}{2}} ,
( \epsilon^f_2 \epsilon^h_1 )^{\frac{1}{2}}
\lesssim 4 \times 10^{-3} &  ( {\rm Re} K )
\\
( \epsilon^f_1 \epsilon^h_3 )^{\frac{1}{2}} ,
( \epsilon^f_3 \epsilon^h_1 )^{\frac{1}{2}}
\lesssim 0.01 & ( | B_d | )
\\
( \epsilon^f_2 \epsilon^h_3 )^{\frac{1}{2}} ,
( \epsilon^f_3 \epsilon^h_2 )^{\frac{1}{2}}
\lesssim 0.03 & ( | B_s | ) ,
\end{array} \ \ \ \ \
\label{eqn:efehLmSmb}
\\
&{}& \Delta {\cal Z}_{\cal D}{\mbox{(b)}} :
\begin{array}{ll}
( \epsilon^h_1 \epsilon^h_2 )^{\frac{1}{2}}
\lesssim 4 \times 10^{-3} / \delta_{12}^{\frac{1}{4}} & ( {\rm Im} K )
\\
( \epsilon^h_1 \epsilon^h_2 )^{\frac{1}{2}} \lesssim 0.02 & ( {\rm Re} K )
\\
( \epsilon^h_1 \epsilon^h_3 )^{\frac{1}{2}} \lesssim 0.03 & ( | B_d | )
\\
( \epsilon^h_2 \epsilon^h_3 )^{\frac{1}{2}} \lesssim 0.09 & ( | B_s | ) .
\end{array}
\label{eqn:ehDZb}
\end{eqnarray}
%%%%%
Here, the constraints for the basis (a) in Eq. (\ref{eqn:efLmSma})
are reproduced roughly from those for the basis (b)
in Eq. (\ref{eqn:efehLmSmb}) under the relation in Eq. (\ref{eqn:(a)-(b)}).
Constraints on the $ u $-$ U $ mixing are estimated
in the bases (a) and (b) as
%%%%%
\begin{eqnarray}
&{}& \Lambda_{\cal U}^{S_-}{\mbox{(a)}} :
\begin{array}{ll}
( \epsilon^f_1 \epsilon^f_2 )^{\frac{1}{2}} \lesssim 0.2 & ( | D^0 | ) ,
\end{array}
\\
&{}& \Lambda_{\cal U}^{S_-}{\mbox{(b)}} :
\begin{array}{ll}
( \epsilon^f_1 \epsilon^h_2 )^{\frac{1}{2}},
( \epsilon^f_2 \epsilon^h_1 )^{\frac{1}{2}}
\lesssim 8 \times 10^{-3} & ( | D^0 | ) ,
\end{array}
\\
&{}& \Delta {\cal Z}_{\cal U}{\mbox{(b)}} :
\begin{array}{ll}
( \epsilon^h_1 \epsilon^h_2 )^{\frac{1}{2}} \lesssim 0.01
& ( | D^0 | ) .
\end{array}
\end{eqnarray}
%%%%%

In supplement to the above constraints
on the $ q $-$ Q $ mixing parameters from the FCNC's,
it is also relevant to consider
the constraints from the flavor-diagonal $ Z $-boson couplings
\cite{LL,qQ2,qQ5}.
The observed branching ratios of the decays
$ Z \rightarrow q_i {\bar q}_i $
imply that the deviations of the flavor-diagonal $ Z $-boson couplings
from the standard model values should be small enough.
Specifically, in the basis (b) with Eq. (\ref{eqn:DZQ-b})
constraints on the $ q $-$ Q $ mixing may be placed roughly as
%%%%%
\begin{eqnarray}
\Delta {\cal Z}_{\cal Q}{\mbox{(b)}} &:&
\epsilon^h_i \lesssim 0.03 \leftarrow
| \Delta {\cal Z}_{\cal Q} [q^\dagger q]_{ii} | \lesssim 10^{-3} .
\label{eqn:ehDZbii}
\end{eqnarray}
%%%%%
On the other hand, in the basis (a)
$ \Delta {\cal Z}_{\cal Q} [q^\dagger q]_{ii} $ of Eq. (\ref{eqn:DZQ-a})
is safely suppressed by $ ( m_{q_i} / m_Q )^2 $ except for $ q_i = t $.

To be more quantitative, we present the results
of detailed numerical calculations
for the $ d $-$ D $ mixing effects in the down-type quark sector
with one singlet $ D $ quark ($ N_D = 1 $ and $ N_U = 0 $)
as a typical case.
We take various values for the model parameters in a reasonable range as
%%%%%
\begin{eqnarray}
&{}&
v = 246 {\rm GeV} , \ v_S = 500 {\rm GeV} ,
\nonumber \\
&{}&
\lambda_{d_i} = \lambda_{d_i}^{(0)} = {\sqrt 2} m_{d_i} / v \
({\rm preliminary}) ,
\nonumber \\
&{}&
| \lambda_D | , | \lambda_D^\prime |
\in [ 0.3 , 1.0 ] \sim m_D / v_S ,
\nonumber \\
&{}&
( v / v_S ) | ( h_d )_i | / | \lambda_D^+ |
\in [ 0 , 0.05 ] \sim \epsilon^h_i ,
\nonumber \\
&{}&
| ( f_D )_i | / | \lambda_D^+ | ,
| ( f_D^\prime )_i | / | \lambda_D^+ |
\in [ 0 , 2.0 ] \sim \epsilon^f_i ,
\nonumber \\
&{}&
{\rm arg} [ h_d , f_D , f_D^\prime , \lambda_D , \lambda_D^\prime ]
\in [ - 0.3 \pi , 0.3 \pi ] .
\nonumber
\end{eqnarray}
%%%%%
Here, the VEV $ v_S $ of the singlet Higgs $ S $ is fixed
to a typical value for definiteness.
The complex phases of the Yukawa couplings
$ h_d , f_D , f_D^\prime , \lambda_D , \lambda_D^\prime $
contribute to the $ CP $ violation in the FCNC's
such as $ \delta_{12} $ for $ \epsilon_K $
of the $ K^0 $-$ {\bar K^0} $ mixing.
The total quark mass matrix $ {\cal M}_{\cal D} $
($ 4 \times 4 $ for $ N_D = 1 $) in Eq. (\ref{eqn:total-MQ})
is given for a set of values of the model parameters.
This preliminary $ {\cal M}_{\cal D} $
with $ \lambda_{d_i} = \lambda_{d_i}^{(0)} $
is diagonalized to evaluate the eigenvalues $ m_{d_i}^{(0)} $
for the ordinary quark masses.
Then, by considering the ratios $ m_{d_i}^{(0)} / m_{d_i} \sim 1 $
we adjust $ \lambda_{d_i} $ to obtain the actual quark masses $ m_{d_i} $:
%%%%%
\begin{eqnarray}
\lambda_{d_i} \rightarrow m_d = 5 {\rm MeV} , \ m_s = 110 {\rm MeV} , \
m_b = 4.2 {\rm GeV} .
\nonumber
\end{eqnarray}
%%%%%
The singlet quark mass $ m_D $ is obtained for the above range
of the model parameters as
%%%%%
\begin{eqnarray}
m_D \sim 100 {\rm GeV} - 1 {\rm TeV} ( v_S = 500 {\rm GeV} ) .
\nonumber
\end{eqnarray}
%%%%%
At the same time, the unitary transformations
$ {\cal V}_{{\cal D}_{\rm L}} $ and $ {\cal V}_{{\cal D}_{\rm R}} $
to specify the quark mass eigenstates are calculated.
The actual CKM matrix $ V $ is reproduced
by adjusting the original unitary matrix $ V_0 $
as $ V_0 = V V_{d_{\rm L}}^{-1} \simeq V V_{d_{\rm L}}^\dagger $
($ V_{u_{\rm L}} = {\bf 1} $ for $ N_U = 0 $):
%%%%%
\begin{eqnarray}
V_0 \rightarrow V ({\rm CKM}) .
\nonumber
\end{eqnarray}
%%%%%
As long as the $ q $-$ Q $ mixing is small enough
to satisfy the constraints from the $ \Delta F = 2 $ meson mixings,
the unitarity violation of the CKM matrix is safely suppressed.

By using these results on the quark masses and $ q $-$ Q $ mixings,
we evaluate the couplings of the quarks
with the gauge bosons and Higgs particles.
Then, the contributions of the $ d $-$ D $ mixing induced FCNC's
to the effective Hamiltonian $ {\cal H}^{\Delta F = 2}_{\rm eff} $
for the $ K^0 $, $ B_d $ and $ B_s $ mixings
are evaluated with Eqs. (\ref{eqn:C1}) -- (\ref{eqn:C4}).
They are compared with the experimental bounds
presented in Table 4 of Ref. \cite{UTfit}
to find the allowed range of the $ d $-$ D $ mixing parameters
$ \epsilon^f_i $ and $ \epsilon^h_i $.
In this analysis, the masses of the Higgs particles are taken typically as
%%%%%
\begin{eqnarray}
m_H = 120 {\rm GeV} , \ m_{S_+} = m_{S_-} = 300 {\rm GeV} .
\end{eqnarray}
%%%%%
Note here that the contributions of the $ S_- $ coupling
in Eqs. (\ref{eqn:C2}) -- (\ref{eqn:C4})
are proportional to $ 1 / m_{S_-}^2 $.
Hence, as $ m_{S_-} $ is larger,
the allowed range of the $ q $-$ Q $ mixing parameters is extended further.

We have made the above calculations
for many samples of the model parameter values.
We show some characteristic results in the following.
The portions of the $ d $-$ D $ mixing effects
in the $ \Delta F = 2 $ meson mixings
for the experimental bounds $ | C_M^k |_{\rm max} $
\cite{UTfit} are denoted by
%%%%%
\begin{eqnarray}
r_k (M) \equiv | C_M^k | / | C_M^k|_{\rm max} .
\label{eqn:rkM}
\end{eqnarray}
%%%%%

%%%%%
\begin{figure}
\begin{center}
\scalebox{.5}{\includegraphics*[0.5cm,1cm][15cm,14cm]{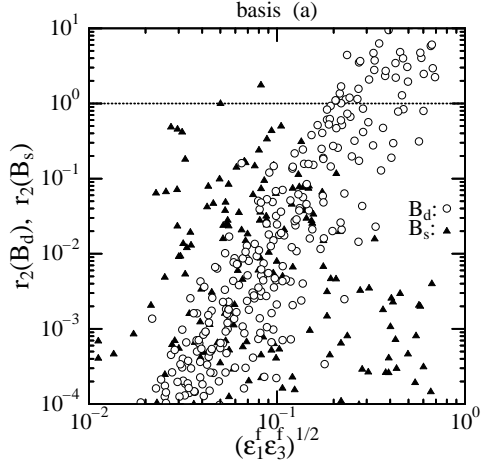}}
\caption{Scatter plots
of $ r_2 ( B_d ) \equiv | C_{B_d}^2 | / | C_{B_d}^2 |_{\rm max} $
($ \circ $)
and $ r_2 ( B_s ) \equiv | C_{B_s}^2 | / | C_{B_s}^2 |_{\rm max} $
($ \blacktriangle $) versus $ ( \epsilon^f_1 \epsilon^f_3 )^{1/2} $
for the $ B_d $-$ {\bar B}_d $ and $ B_s $-$ {\bar B}_s $ mixings,
respectively, which are provided by the $ S_- $ coupling
$ \Lambda_{\cal D}^{S_-} [d^c d] $ in the basis (a).}
\label{efCk-a1}
\end{center}
\end{figure}
%%%%%
\begin{figure}
\begin{center}
\scalebox{.5}{\includegraphics*[0.5cm,1cm][15cm,14cm]{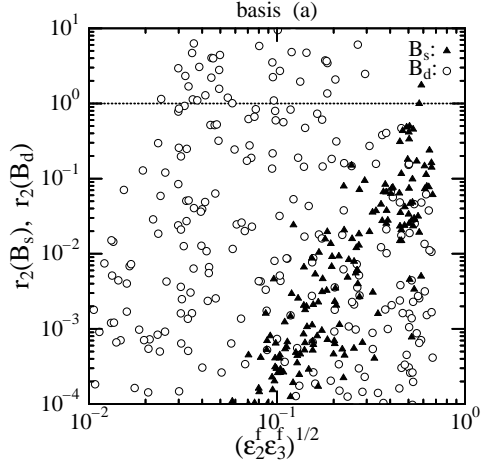}}
\caption{Scatter plots
of $ r_2 ( B_s ) \equiv | C_{B_s}^2 | / | C_{B_s}^2 |_{\rm max} $
($ \blacktriangle $)
and $ r_2 ( B_d ) \equiv | C_{B_d}^2 | / | C_{B_d}^2 |_{\rm max} $
($ \circ $) versus $ ( \epsilon^f_2 \epsilon^f_3 )^{1/2} $,
similarly to Fig. \ref{efCk-a1}.}
\label{efCk-a2}
\end{center}
\end{figure}
%%%%%
%%%%%
\begin{figure}
\begin{center}
\scalebox{.5}{\includegraphics*[0.5cm,1cm][15cm,14cm]{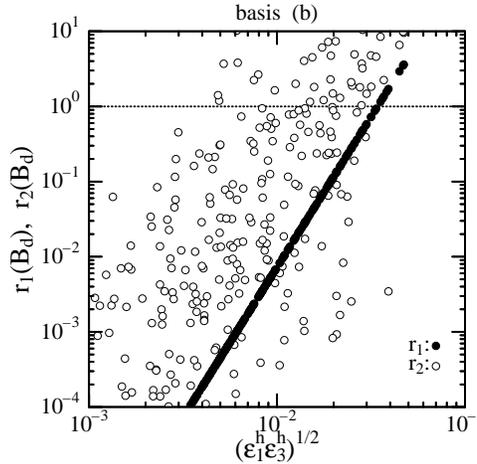}}
\caption{Scatter plots
of $ r_1 ( B_d ) \equiv | C_{B_d}^1 | / | C_{B_d}^1 |_{\rm max} $
($ \bullet $)
and $ r_2 ( B_d ) \equiv | C_{B_d}^2 | / | C_{B_d}^2 |_{\rm max} $
($ \circ $) versus $ ( \epsilon^h_1 \epsilon^h_3 )^{1/2} $
for the $ B_d $-$ {\bar B}_d $ mixing,
which are provided by the $ Z $ coupling
$ \Delta {\cal Z}_{\cal D} [d^\dagger d] $
and the $ S_- $ coupling $ \Lambda_{\cal D}^{S_-} [d^c d] $,
respectively, in the basis (b).}
\label{ehCk-b1}
\end{center}
\end{figure}
%%%%%
\begin{figure}
\begin{center}
\scalebox{.5}{\includegraphics*[0.5cm,1cm][15cm,14cm]{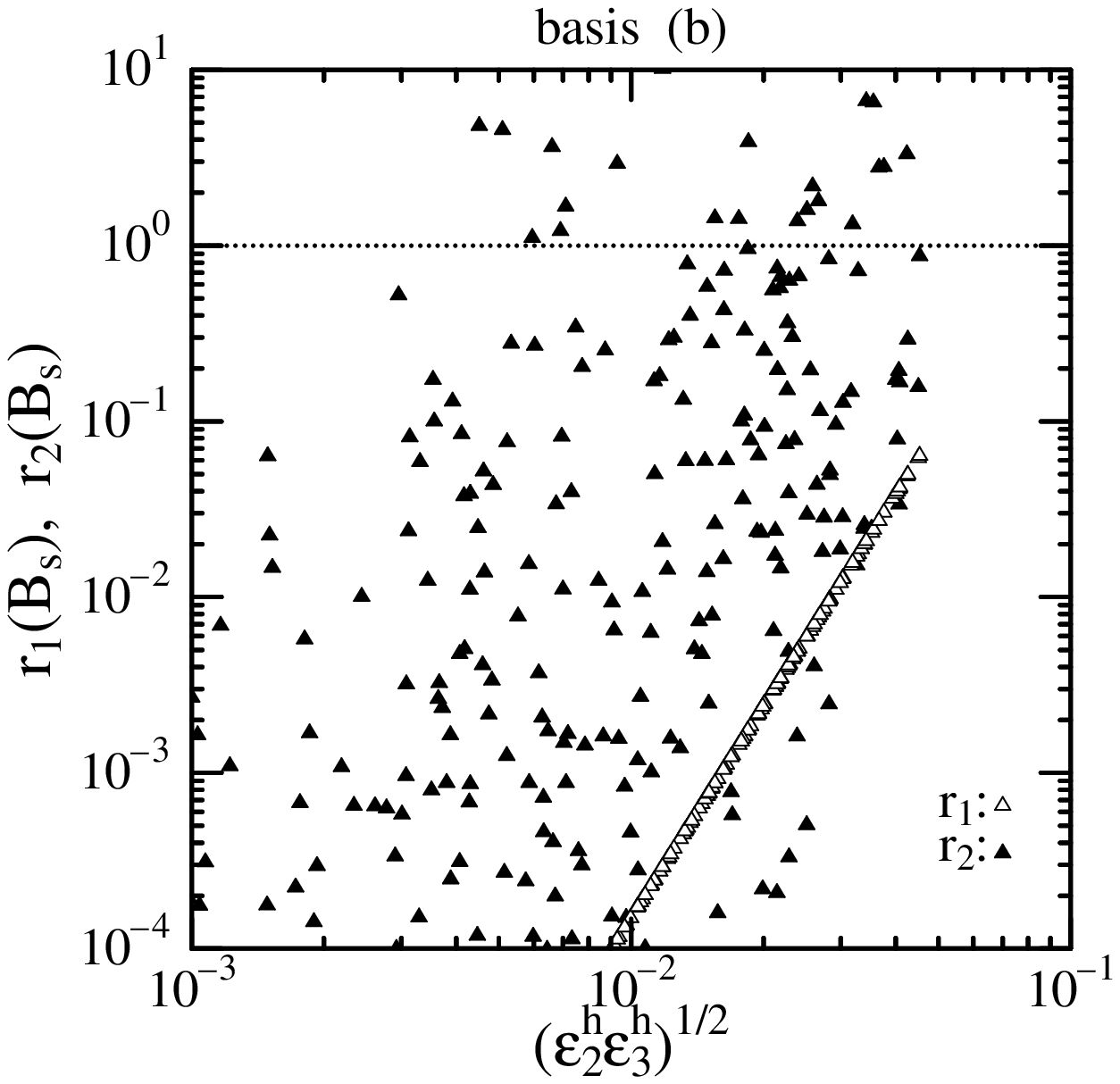}}
\caption{Scatter plots
of $ r_1 ( B_s ) \equiv | C_{B_s}^1 | / | C_{B_s}^1 |_{\rm max} $
($ \vartriangle $)
and $ r_2 ( B_s ) \equiv | C_{B_s}^2 | / | C_{B_s}^2 |_{\rm max}$
($ \blacktriangle $) versus $ ( \epsilon^h_2 \epsilon^h_3 )^{1/2} $
for the $ B_s $-$ {\bar B}_s $ mixing,
similarly to Fig. \ref{ehCk-b1}.}
\label{ehCk-b2}
\end{center}
\end{figure}
%%%%%

In Fig. \ref{efCk-a1} scatter plots
of $ r_2 ( B_d ) $ ($ \circ $) and $ r_2 ( B_s ) $ ($ \blacktriangle $)
versus $ ( \epsilon^f_1 \epsilon^f_3 )^{1/2} $ are shown
for the $ B_d $-$ {\bar B}_d $ and $ B_s $-$ {\bar B}_s $ mixings,
respectively, which are provided by the singlet $ S_- $ coupling
$ \Lambda_{\cal D}^{S_-} [d^c d] $ in the basis (a).
Similar plots of $ r_2 ( B_s ) $ ($ \blacktriangle $)
and $ r_2 ( B_d ) $ ($ \circ $)
versus $ ( \epsilon^f_2 \epsilon^f_3 )^{1/2} $
are shown in Fig. \ref{efCk-a2}.
The bounds for the $ K^0 $-$ {\bar K}^0 $ mixing have been checked
to be satisfied in these plots.
The results in Figs. \ref{efCk-a1} and \ref{efCk-a2}
are in accordance with the rough estimates
to obtain the bounds in Eq. (\ref{eqn:efLmSma}).
The dominant effects of the $ S_- $ coupling
$ \Lambda_{\cal D}^{S_-} [d^c d] $ of Eq. (\ref{eqn:LmSma})
are estimated in Eq. (\ref{eqn:C2})
as $ | C_{B_d}^2 |
\sim ( m_b / v_S )^2 ( \epsilon^f_1 \epsilon^f_3 )^2 / m_{S_-}^2 $
and $ | C_{B_s}^2 |
\sim ( m_b / v_S )^2 ( \epsilon^f_2 \epsilon^f_3 )^2 / m_{S_-}^2 $.
In the log-log plot of Fig. \ref{efCk-a1},
$ r_2 (B_d) $ ($ \circ $) shows roughly the expected linear correlation
with $ ( \epsilon^f_1 \epsilon^f_3 )^{\frac{1}{2}} $,
while $ r_2 (B_s) $ ($ \blacktriangle $) are distributed independently
of $ ( \epsilon^f_1 \epsilon^f_3 )^{\frac{1}{2}} $.
We see the similar feature in Fig. \ref{efCk-a2}
for $ r_2 (B_s) $ ($ \blacktriangle $) and $ r_2 (B_d) $ ($ \circ $)
versus $ ( \epsilon^f_2 \epsilon^f_3 )^{\frac{1}{2}} $.
Precisely, in the basis (a)
the $ d $-$ D $ mixing parameters $ \epsilon^f_i $
are defined with the $ f_D^+ $ coupling,
while the singlet $ S_- $ coupling $ \Lambda_{\cal D}^{S_-} $
is given by the $ f_D^- $ coupling
($ | f_D^+ | \sim | f_D^- | $ generally).
This fact provides the appreciable spreads
in the plots of $ r_2 (B_d) $
versus $ ( \epsilon^f_1 \epsilon^f_3 )^{\frac{1}{2}} $
and $ r_2 (B_s) $
versus $ ( \epsilon^f_2 \epsilon^f_3 )^{\frac{1}{2}} $.
We find in these plots that the $ d $-$ D $ mixing parameters
$ ( \epsilon^f_1 \epsilon^f_3 )^{\frac{1}{2}} $
and $ ( \epsilon^f_2 \epsilon^f_3 )^{\frac{1}{2}} $
are really constrained for $ r_2 (B_d) \leq 1 $ and $ r_2 (B_s) \leq 1 $,
respectively, as shown in Eq. (\ref{eqn:efLmSma}).
We note particularly that 
both the bounds for the $ B_d $-$ {\bar B}_d $ and $ B_s $-$ {\bar B}_s $
mixings may be saturated simultaneously with $ \epsilon^f_3 \sim 1 $
and $ ( \epsilon^f_1 \epsilon^f_2 )^{\frac{1}{2}} \sim 0.1 $
without conflicting with the bonds for the $ K^0 $-$ {\bar K}^0 $ mixing.
Generally, in the basis (a) the right-handed $ d $-$ D $ mixing
is rather tolerable with $ \epsilon^f_i \sim 0.1 - 1 $.
This is because the right-handed components
of the ordinary and singlet quarks are indistinguishable
with respect to the gauge interactions.

In Fig. \ref{ehCk-b1} scatter plots
of $ r_1 ( B_d ) $ ($ \bullet $) and $ r_2 ( B_d ) $ ($ \circ $)
versus $ ( \epsilon^h_1 \epsilon^h_3 )^{1/2} $
are shown for the $ B_d $-$ {\bar B}_d $ mixing,
which are provided by the $ Z $-boson coupling
$ \Delta {\cal Z}_{\cal D} [d^\dagger d] $
and the singlet $ S_- $ coupling $ \Lambda_{\cal D}^{S_-} [d^c d] $,
respectively, in the basis (b).
Similar plots of $ r_1 ( B_s ) $ ($ \vartriangle $)
and $ r_2 ( B_s ) $ ($ \blacktriangle $)
versus $ ( \epsilon^h_2 \epsilon^h_3 )^{1/2} $
are shown in Fig. \ref{ehCk-b2}
for the $ B_s $-$ {\bar B}_s $ mixing.
The bounds for the $ K^0 $-$ {\bar K}^0 $ mixing
have been checked to be satisfied in these plots.
The flavor-diagonal $ Z $-boson couplings
have also been checked to satisfy
$ | \Delta {\cal Z}_{\cal Q} [q^\dagger q]_{ii} | < 3 \times 10^{-3} $,
as considered in Eq. (\ref{eqn:ehDZbii}).
The effects of the $ Z $-boson coupling
$ \Delta {\cal Z}_{\cal Q} [q^\dagger q] $ of Eq. (\ref{eqn:DZQ-b})
are estimated in Eq. (\ref{eqn:C1}) for the basis (b)
as $ | C_{B_d}^1 | \sim ( \epsilon^h_1 \epsilon^h_3 )^2 {\sqrt 2} G_F $
and $ | C_{B_s}^1 | \sim ( \epsilon^h_2 \epsilon^h_3 )^2 {\sqrt 2} G_F $.
In the log-log plot of Fig. \ref{ehCk-b1},
$ r_1 (B_d) $ ($ \bullet $) shows clearly the linear correlation
with $ ( \epsilon^h_1 \epsilon^h_3 )^{\frac{1}{2}} $
[here more precisely
$ ( \epsilon_{q_{\rm L}} )_i \simeq ( \epsilon_{q_{\rm L}}^\prime )_i
\simeq \epsilon^h_i $ for the left-handed $ q $-$ Q $ mixing].
This is also the case in Fig. \ref{ehCk-b2}
for $ r_1 (B_s) $ ($ \vartriangle $)
versus $ ( \epsilon^h_2 \epsilon^h_3 )^{\frac{1}{2}} $.
The $ d $-$ D $ mixing
with $ \epsilon^h_1 \sim 0.03 $ and $ \epsilon^h_3 \sim 0.03 $
may saturate the bound for the $ B_d $-$ {\bar B}_d $ mixing
via the $ Z $-boson coupling,
$ r_1 ( B_d ) \equiv| C_{B_d}^1 | / | C_{B_d}^1 |_{\rm max} \approx 1 $,
as seen in Fig. \ref{ehCk-b1},
which also provides significant effects $ \sim 0.1 \% $
on the flavor-diagonal $ Z $-boson couplings in Eq. (\ref{eqn:ehDZbii}).
On the other hand, as long as $ \epsilon^h_i \lesssim 0.03 $,
the $ d $-$ D $ mixing effect $ C_{B_s}^1 $
via the $ Z $-boson coupling is fairly below
the bound for the $ B_s $-$ {\bar B}_s $ mixing,
$ r_1 ( B_s ) \equiv | C_{B_s}^1 | / | C_{B_s}^1 |_{\rm max} < 0.1 $,
as seen in Fig. \ref{ehCk-b2}.
It should be noted here that
as seen in Figs. \ref{ehCk-b1} and \ref{ehCk-b2},
the contributions
$ C_{B_d}^2 $ ($ \circ $) and $ C_{B_s}^2 $ ($ \blacktriangle $)
via the singlet $ S_- $ coupling
$ \Lambda_{\cal D}^{S_-} [d^c d] $ of Eq. (\ref{eqn:LmSmb})
may dominate over the $ Z $-boson coupling effects
in the parameter region of $ \epsilon^f_i > \epsilon^h_i $,
where the bounds in Eq. (\ref{eqn:efehLmSmb}) are applicable.
In particular, the case of $ \epsilon^f_i \gg \epsilon^h_i $
is connected gradually to a suitable parameter region in the basis (a).

In short, remarkable effects may be provided particularly
for the $ B $ mesons through the significant mixing
between the $ b $ quark and the singlet $ D $ quark
with $ \epsilon^f_3 \sim 1 $ or $ \epsilon^h_3 \sim 0.03 $,
as seen in the above.
They are fairly expected to serve as new physics
for the flavor-changing processes
and $ CP $-violation in the $ B $ meson physics.
Specifically, it is well known that
there is tension between the experimental constraints
and the standard model contribution
to the $ b \rightarrow s \gamma $ process,
and hence this process should not be used as a constraint at present.
The recent constraint by HFAG \cite{HFAG} is given
as the average of the data of BABAR, Belle, and CLEO,
$ {\rm Br} ( b \rightarrow s \gamma )
= ( 352 \pm 23 \pm 9 ) \times 10^{-6} $,
while the recent predictions of the standard model contribution
are given as $ {\rm Br} ( b \rightarrow s \gamma )
= ( 315 \pm 23 ) \times 10^{-6} $ \cite{smbsgamma1}
and $ {\rm Br} ( b \rightarrow s \gamma )
= ( 298 \pm 26 ) \times 10^{-6} $ \cite{smbsgamma2}.
Even though this discrepancy is small,
it may be confirmed by future experiments.
The FCNC's via the $ d $-$ D $ mixing
can provide a solution of the discrepancy.
This topic is, however, beyond the scope of the present work,
and will be studied elsewhere.

%%%%%%%%%%
\section{Decays of singlet quarks and Higgs particles}
\label{sec:Qprocesses}
%%%%%%%%%%

We now investigate the decays of the singlet quarks and Higgs particles,
which will provide distinct signatures upon their productions at the LHC.
The flavor-changing interactions
between the singlet quarks $ Q $ and the ordinary quarks $ q $
are relevant for these decays at the tree level.
Specifically, the left-handed $ Z $-boson couplings are given as
%%%%%
\begin{eqnarray}
{\cal Z}_{\cal Q} [q^\dagger Q]_{ia}
= {\cal Z}_{\cal Q} [Q^\dagger q]_{ai}^*
= ( V_{q_{\rm L}}^\dagger \epsilon_{q_{\rm L}} )_{ia}
\simeq ( \epsilon_{q_{\rm L}} )_{ia} ,
\label{eqn:ZqQ}
\end{eqnarray}
%%%%%
while the right-handed ones are absent.
Note here that $ {\cal Z}_{\cal Q} [q^\dagger Q]
= \Delta {\cal Z}_{\cal Q} [q^\dagger Q] $,
as shown in Eq. (\ref{eqn:ZqQ-A});
the $ q $-$ Q $ transitions in the $ Z $-boson couplings
are just induced as the $ q $-$ Q $ mixing effect.
The left-handed $ W $-boson couplings are given
in terms of the $ Z $-boson couplings and the CKM matrix
in a good approximation up to the second order of the small
$ q $-$ Q $ mixing as
%%%%%
\begin{eqnarray}
{\cal V} [u^\dagger D]_{ia} &=& {\cal V} [D^\dagger u]_{ai}^*
\simeq ( V {\cal Z}_{\cal D} [d^\dagger D] )_{ia} ,
\label{eqn:VuD}
\\
{\cal V} [d^\dagger U]_{ia} &=& {\cal V} [U^\dagger d]_{ai}^*
\simeq ( V^\dagger {\cal Z}_{\cal U} [u^\dagger U] )_{ia} ,
\label{eqn:VdU}
\end{eqnarray}
%%%%%
while the right-handed ones are absent.
The neutral scalar couplings are given as
%%%%%
\begin{eqnarray}
\Lambda_{\cal Q}^H [q^c Q]_{ia}
&=& ( m_{q_i} / v ) {\cal Z}_{\cal Q} [q^\dagger Q]_{ia} ,
\label{eqn:LHqQ}
\\
\Lambda_{\cal Q}^H [Q^c q]_{ai}
&=& ( m_{Q_a} / v ) {\cal Z}_{\cal Q} [Q^\dagger q]_{ai} ,
\label{eqn:LHQq}
\\
\Lambda_{\cal Q}^{S_+} [q^c Q]_{ia}
&=& - ( m_{q_i} / v_S ) {\cal Z}_{\cal Q} [q^\dagger Q]_{ia} ,
\label{eqn:LS+qQ}
\\
\Lambda_{\cal Q}^{S_+} [Q^c q]_{ai}
&=& - ( m_{Q_a} / v_S ) {\cal Z}_{\cal Q} [Q^\dagger q]_{ai} ,
\label{eqn:LS+Qq}
\\
\Lambda_{\cal Q}^{S_-} [q^c Q]_{ia}
&=& ( V_{q_{\rm R}}^\dagger f_Q^- V_{Q_{\rm L}}
- \epsilon_{q_{\rm R}}^\prime \lambda_Q^- V_{Q_{\rm L}} )_{ia}/{\sqrt 2} ,
\label{eqn:LS-qQ}
\\
\Lambda_{\cal Q}^{S_-} [Q^c q]_{ai}
&=& ( - \epsilon_{q_{\rm R}}^\dagger f_Q^-
\epsilon_{q_{\rm L}}^{\prime \dagger}
- V_{Q_{\rm R}}^\dagger \lambda_Q^-
\epsilon_{q_{\rm L}}^{\prime \dagger} )_{ai}/{\sqrt 2} .
\label{eqn:LS-Qq}
\end{eqnarray}
%%%%%
Here, $ \Lambda_{\cal Q}^{\phi^0_r} [q^c Q]_{ia} $
stands for $ {\bar q}_{i{\rm R}} Q_{a{\rm L}} \phi^0_r $,
and $ \Lambda_{\cal Q}^{\phi^0_r} [Q^c q]_{ai} $
for $ {\bar Q}_{a{\rm R}} q_{i{\rm L}} \phi^0_r $, respectively,
in terms of the Dirac fields.
The relations among the gauge and scalar couplings
in Eqs. (\ref{eqn:ZqQ}) -- (\ref{eqn:LS+Qq})
are derived in the Appendix \ref{sec:couplings}.

%%%%%
\subsection{Singlet quark decays}
\label{subsec:Q-decays}
%%%%%

We first investigate the singlet quark decays.
We describe the essential features
by considering the case of one down-type singlet quark $ D $
($ a = 1 $ is omitted for $ N_D = 1 $ and $ N_U = 0 $).
Similar results are obtained
in the general cases of some $ D $ and $ U $ quarks,
especially for the lightest singlet quark.
While the heavier singlet quarks may decay dominantly
into the lighter singlet quarks and Higgs particles in the general cases,
we concentrate on the decays of the lightest singlet quark
producing the ordinary quarks.

The partial widths of the relevant decay modes are calculated
(when they are kinematically allowed) as
%%%%%
\begin{eqnarray}
\Gamma ( D \rightarrow u_i W )
&=& \frac{G_F}{\sqrt 2} \frac{m_D^3}{8 \pi} g( x_W , x_{u_i} )
| {\cal V} [u^\dagger D]_i |^2 ,
\label{eqn:DuW}
\\
\Gamma ( D \rightarrow d_i Z )
&=& \frac{G_F}{\sqrt 2} \frac{m_D^3}{16 \pi} ( 1 - 3 x_Z^4 + 2 x_Z^6 )
\nonumber \\
&{}& \times | {\cal Z}_{\cal D} [d^\dagger D]_i |^2 ,
\label{eqn:DdZ}
\\
\Gamma ( D \rightarrow d_i H )
&=& \frac{m_D}{16 \pi} ( 1 - x_H^2 )^2
\nonumber \\
&{}& \times ( | \Lambda_{\cal D}^H [D^c d]_i |^2
+ | \Lambda_{\cal D}^H [d^c D]_i |^2 ) ,
\label{eqn:DdH}
\\
\Gamma ( D \rightarrow d_i S_\pm )
&=& \frac{m_D}{16 \pi} ( 1 - x_{S_\pm}^2 )^2
\nonumber \\
&{}& \times ( | \Lambda_{\cal D}^{S_\pm} [D^c d]_i |^2
+ | \Lambda_{\cal D}^{S_\pm} [d^c D]_i |^2 ) , \ \ \
\label{eqn:DdS}
\end{eqnarray}
%%%%%
where $ x_W = m_W / m_D $, $ x_Z = m_Z / m_D $, $ x_H = m_H / m_D $,
$ x_{S_\pm} = m_{S_\pm} / m_D $, $ x_{u_i} = m_{u_i} / m_D $,
and
%%%%%
\begin{eqnarray}
g(x,y) &=& [ 1 - ( x + y )^2 ]^{1/2} [ 1 - ( x - y )^2 ]^{1/2}
\nonumber \\
&{}& \times  [ x^2 ( 1 - 2 x^2 + y^2 ) + ( 1 - y^2 )^2 ] .
\end{eqnarray}
%%%%%
The scalar mixing is neglected ($ O_{\phi} = {\bf 1} $) for a while.
The kinematic effects of $ m_{d_i} / m_D \lesssim 0.02 $
for $ m_D \gtrsim 200 {\rm GeV} $ may be neglected in a good approximation,
while $ \Gamma ( D \rightarrow t W ) $ depends sensibly on $ m_t / m_D $.

The flavor-structure of the $ d $-$ D $ mixing is measured
manifestly in the $ D $ decays into the ordinary quarks
$ d_i = d , s , b $ and the $ Z $ boson as
%%%%%
\begin{eqnarray}
\Gamma ( D \rightarrow d_i Z ) \propto | ( \epsilon_{d_{\rm L}} )_i |^2 ,
\end{eqnarray}
%%%%%
where $ {\cal Z}_{\cal D} [d^\dagger D]_i
\simeq ( \epsilon_{d_{\rm L}} )_i $ for the $ Z $-boson coupling.
The partial width of the $ D $ quark decays
producing the $ Z $ boson is inclusively estimated for the reference as
%%%%%
\begin{eqnarray}
\Gamma_D (Z)
& \equiv &
\sum_i \Gamma ( D \rightarrow d_i Z )
\nonumber \\
& \sim & 20 {\rm MeV} \times \left( \frac{m_D}{500 {\rm GeV}} \right)^3
\frac{| \epsilon^h |^2}{( 0.03 )^2} ,
\label{eqn:GDZ}
\end{eqnarray}
%%%%%
where $ ( \epsilon_{d_{\rm L}} )_i \sim \epsilon^h_i $
with $ | \epsilon^h | \equiv
[ ( \epsilon^h_1 )^2 + ( \epsilon^h_2 )^2 + ( \epsilon^h_3 )^2 ]^{1/2} $
in the basis (b),
and $ \epsilon^h_i \rightarrow ( m_{d_i} / m_D ) \epsilon^f_i $
for $ ( \epsilon_{d_{\rm L}} )_i $ in the basis (a).
By considering Eqs. (\ref{eqn:ZqQ}), (\ref{eqn:VuD}),
(\ref{eqn:LHqQ}) and (\ref{eqn:LHQq})
with
$ | \Lambda_{\cal D}^H [d^c D]_i |^2 / | \Lambda_{\cal D}^H [D^c d]_i |^2
= ( m_{d_i} / m_D )^2 \ll 1 $,
we find the well-known relations \cite{QqH-1}
%%%%%
\begin{eqnarray}
\Gamma ( D \rightarrow u_i W )
& \sim & 2 \Gamma ( D \rightarrow d_i Z ) ,
\label{eqn:GD-WZ}
\\
\Gamma ( D \rightarrow d_i H )
& \sim & \Gamma ( D \rightarrow d_i Z ) ,
\label{eqn:GD-HZ}
\end{eqnarray}
%%%%%
or inclusively
%%%%%
\begin{eqnarray}
\Gamma_D (W)
& \equiv &
\sum_i \Gamma ( D \rightarrow u_i W ) \sim 2 \Gamma_D (Z) ,
\label{eqn:GDW}
\\
\Gamma_D (H)
& \equiv &
\sum_i \Gamma ( D \rightarrow d_i H ) \sim \Gamma_D (Z) ,
\label{eqn:GDH}
\end{eqnarray}
%%%%%
where $ O_\phi = {\bf 1} $.
The actual values of these widths are evaluated
depending on the kinematic factors and the CKM mixing.

In the present model with the complex singlet Higgs,
the decays of the singlet quark $ D $
producing the singlet Higgs scalars $ S_\pm $
are possible for $ m_D > m_{S_\pm} + m_{d_i} $.
Especially, it is remarkable that the decays with $ S_- $
may dominate over the other decay modes.
By considering
$ ( f_D^- )_i \sim ( \epsilon_{d_{\rm R}}^\prime )_i \lambda_D^-
\sim ( m_D / v_S ) \epsilon^f_i $
and $ \lambda_D^- ( \epsilon_{d_{\rm L}}^{\prime \dagger} )_i
\sim ( m_D / v_S ) \epsilon^h_i $
in Eqs. (\ref{eqn:LS-qQ}) and (\ref{eqn:LS-Qq}),
the $ S_- $ couplings are estimated roughly as
%%%%%
\begin{eqnarray}
\Lambda_{\cal D}^{S_-} [d^c D]_i & \sim & ( m_D / v_S ) \epsilon^f_i ,
\\
\Lambda_{\cal D}^{S_-} [D^c d]_i & \sim & ( m_D / v_S ) \epsilon^h_i .
\end{eqnarray}
%%%%%
Here, for the sake of convenience
we adopt $ \epsilon^h_i = ( m_{d_i} / m_D ) \epsilon^f_i $
in the basis (a) though $ h_d = {\bf 0} $,
while $ \epsilon^f_i = ( v_S / m_D ) | 2 ( f_Q )_i |/{\sqrt 2} $
in the basis (b) though $ f_Q^+ = {\bf 0} $,
as discussed concerning Eq. (\ref{eqn:(a)-(b)}).
Then, we estimate roughly the partial width of the $ D $ decays
producing $ S^- $ as
%%%%%
\begin{eqnarray}
\Gamma_D ( S_- ) & \equiv & \sum_i \Gamma ( D \rightarrow d_i S_- )
\nonumber \\
& \sim & 10 {\rm MeV} \times 
\left( \frac{m_D}{500 {\rm GeV}} \right)^3
\left( \frac{500 {\rm GeV}}{v_S} \right)^2
\nonumber \\
&{}& \times \frac{| \epsilon^f |^2 + | \epsilon^h |^2}{( 0.03 )^2} ,
\label{eqn:GDSm}
\end{eqnarray}
%%%%%
where $ | \epsilon^f | \equiv
[ ( \epsilon^f_1 )^2 + ( \epsilon^f_2 )^2 + ( \epsilon^f_3 )^2 ]^{1/2} $.
(The actual value is reduced to some extent
by the kinematic factor for $ m_D \sim m_{S^-} $.)
This width $ \Gamma_D ( S_- ) $ for the decays into the singlet scalar
$ S_- $ dominates over the reference width $ \Gamma_D (Z) $
for the decays into the $ Z $ boson in Eq. (\ref{eqn:GDZ})
for $ | \epsilon^f |^2 \gg | \epsilon^h |^2 $, especially in the basis (a)
with $ \epsilon^h_i \rightarrow ( m_{d_i} / m_D ) \epsilon^f_i $.
As for the $ D $ decays with $ S_+ $,
the partial width is simply related
to $ \Gamma_D (Z) $ by Eq. (\ref{eqn:LS+Qq}) as
%%%%%
\begin{eqnarray}
\Gamma_D ( S_+ ) \equiv \sum_i \Gamma ( D \rightarrow d_i S_+ )
\sim ( v / v_S )^2 \Gamma_D (Z) ,
\label{eqn:GDSp}
\end{eqnarray}
%%%%%
which amounts to $ O ( 10 \% ) $ of $ \Gamma_D (Z) $
for $ v_S \approx 500 {\rm GeV} $.
Even this slight enhancement due to $ \Gamma_D (S_+) $
for the $ D $ decays into the scalars $ H $ and $ S_+ $
might serve as an experimental signature
for the singlet Higgs even if $ S_- $ is absent
in the model with one real $ S \equiv S_+ $.

Here, it should be noted that
the $ S_- $ coupling may even provide significant contributions
to the decays $ D \rightarrow d_i H $
through the $ H $-$ S_- $ mixing $ \epsilon_{H S_-} $ in $ O_\phi $.
In fact, the $ d $-$ D $ coupling with the standard Higgs $ H $
(more precisely the mass eigenstate $ \phi_1 \simeq H $
with $ \epsilon_{H S_-} \ll 1 $) is replaced in Eq. (\ref{eqn:DdH}) as
%%%%%
\begin{equation}
\Lambda_{\cal D}^H
\rightarrow \Lambda_{\cal D}^H + \epsilon_{H S_-} \Lambda_{\cal D}^{S_-} .
\label{eqn:LmHS}
\end{equation}
%%%%%
Then, instead of Eq. (\ref{eqn:GDH}) we obtain
%%%%%
\begin{eqnarray}
\Gamma_D (H) & \sim & \Gamma_D (Z) + \epsilon_{H S_-}^2 \Gamma_D ( S_- ) ,
\label{eqn:GDSmH}
\end{eqnarray}
%%%%%
where the interference term between $ \Lambda_{\cal D}^H $
and $ \Lambda_{\cal D}^{S_-} $ is omitted for simplicity.
This enhancement of $ \Gamma_D (H) $
with $ \epsilon_{H S_-} \Lambda_{\cal D}^{S_-} $ in Eq. (\ref{eqn:LmHS})
is valid even when the decays $ D \rightarrow d_i S_- $
are forbidden kinematically for $ m_D < m_{S_-} + m_{d_i} $.
In the presence of a small but sizable $ H $-$ S_- $ mixing
$ \epsilon_{H S_-} \sim 0.01 - 0.1 $
the singlet quark decays $ D \rightarrow d_i H $
may become the dominant modes, particularly in the basis (a)
due to the substantial suppression of $ D \rightarrow d_i Z $
with $ | {\cal Z}_{\cal D} [d^\dagger D]_i |^2
\sim ( m_{d_i} / m_D )^2 ( \epsilon^f_i )^2 \lesssim 10^{-4} $.
Hence, if $ \Gamma_D (H) \gg \Gamma_D (Z) $ is confirmed experimentally,
which is contrary to the usual expectation of Eq. (\ref{eqn:GDH}),
it will provide a distinct evidence
for the complex singlet Higgs field $ S $
with the $ H $-$ S_- $ mixing.
The decays $ D \rightarrow d_i S_+ $ may also be enhanced substantially
as $ \Gamma_D ( S_+ ) \sim \epsilon_{S_+ S_-}^2 \Gamma_D ( S_- ) $
via a sizable $ S_+ $-$ S_- $ mixing $ \epsilon_{S_+ S_-} $.

We present in the following the detailed estimates
on the widths of the relevant decay modes,
where the constraints on the $ d $-$ D $ mixing
from the $ \Delta F = 2 $ meson mixings
and the diagonal $ Z $-boson couplings are checked
to be satisfied according to the numerical calculations
performed in Sec. \ref{sec:DF2}.

%%%%%
\begin{figure}
\begin{center}
\scalebox{.5}{\includegraphics*[0.5cm,1cm][15cm,14cm]{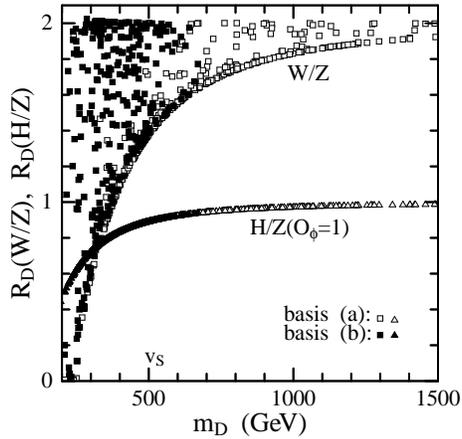}}
\caption{$ R_D (W/Z) \equiv \Gamma_D (W) / \Gamma_D (Z) $ versus $ m_D $
is shown for the bases (a) ($ \Box $) and (b) ($ \blacksquare $).
$ R_D (H/Z) \equiv \Gamma_D (H) / \Gamma_D (Z) $ versus $ m_D $
is also shown for the bases (a) ($ \vartriangle $)
and (b) ($ \blacktriangle $).
Here, $ v_S = 500 {\rm GeV} $,
and the Higgs scalar mixing is assumed to be absent ($ O_\phi = {\bf 1} $).}
\label{mDGWHZ}
\end{center}
\end{figure}
%%%%%
We suitably denote the ratios of the relevant widths
to the reference width as
%%%%%
\begin{eqnarray}
R_D (X/Z) \equiv \frac{\Gamma_D (X)}{\Gamma_D (Z)} ,
\label{eqn:RD-X/Z}
\end{eqnarray}
%%%%%
where $ X = W , H , S_+ , S_- $.
For the usual decay modes
$ D \rightarrow u_i W $, $ D \rightarrow d_i Z $
and $ D \rightarrow d_i H $, scatter plots
of $ R_D (W/Z) $ ($ \Box $, $ \blacksquare $)
and $ R_D (H/Z) $ ($ \vartriangle $, $ \blacktriangle $)
versus the singlet quark mass $ m_D $ are shown in Fig. \ref{mDGWHZ}
for the bases (a) ($ \Box $, $ \vartriangle $)
and (b) ($ \blacksquare $, $ \blacktriangle $).
Here, $ v_S = 500 {\rm GeV} $,
and the Higgs scalar mixing is assumed to be absent ($ O_\phi = {\bf 1} $).
Similar results are obtained for the bases (a) and (b)
since these bases are equivalently related
to each other by the unitary transformation,
as discussed in Sec. \ref{sec:mixing-FCI}.
Note here that larger values may be obtained
for the singlet quark mass $ m_D $ with a given singlet Higgs VEV $ v_S $
in the basis (a) ($ \Box $, $ \vartriangle $),
which is due to the significant contribution
of the $ q $-$ Q $ mixing term $ \Delta_{qQ} = f_Q^+ v_S / {\sqrt 2} $
for $ | \epsilon^f | \sim 1 $.
The lower boundary curve for $ R_D (W/Z) $ reflects the kinematic factor
of the dominant top contribution $ D \rightarrow t W $ with
$ | {\cal V} [u^\dagger D]_3 |^2 \gg | {\cal V} [u^\dagger D]_{1,2} |^2 $.
In this case the singlet $ D $ quark mixes mainly with the $ b $ quark
as $ | ( \epsilon_{d_{\rm L}} )_3 |^2
\gg | ( \epsilon_{d_{\rm L}} )_{1,2} |^2 $.
On the other hand, in the case that the top contribution is negligible with
$ | {\cal V} [u^\dagger D]_3 |^2 \ll | {\cal V} [u^\dagger D]_{1,2} |^2 $,
the asymptotic value $ R_D (W/Z) = 2 $ is almost saturated
for $ m_D \gtrsim 300 {\rm GeV} $.
We also see that $ R_D (H/Z) $ approaches the asymptotic value
$ R_D (H/Z) = 1 $ showing the kinematic dependence on $ m_D $.
These results really confirm the usual expectation
in Eqs. (\ref{eqn:GDW}) and (\ref{eqn:GDH}).
It should, however, be remarked
that as shown in Eq. (\ref{eqn:GDSmH}),
the $ D $ decays with the standard Higgs $ H $
may be enhanced substantially as $ R_D (H/Z) \gg 1 $
due to the singlet Higgs coupling $ \Lambda_{\cal D}^{S_-} $
via the $ H $-$ S_- $ mixing.

\begin{figure}
\begin{center}
\scalebox{.5}{\includegraphics*[0.5cm,1cm][15cm,14cm]{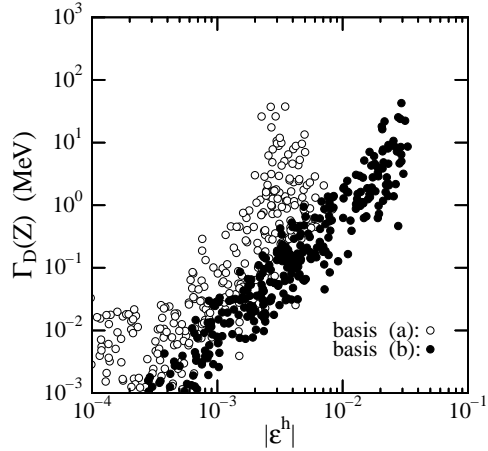}}
\caption{$ \Gamma_D (Z) $ versus $ | \epsilon^h | $
in the bases (a) ($ \circ $) and (b) ($ \bullet $).
Here, $ \epsilon^h_i = ( m_{q_i} / m_D ) \epsilon^f_i $
as Eq. (\ref{eqn:(a)-(b)}) is adopted in the basis (a)
though $ \epsilon^h_i = 0 $ formally.}
\label{ehGZ}
\end{center}
\end{figure}

The reference width $ \Gamma_D (Z) $ versus
the magnitude of the left-handed $ d $-$ D $ mixing
$ | \epsilon^h | $, as given in Eq. (\ref{eqn:GDZ}),
is shown in Fig. \ref{ehGZ}.
Here, the marks $ \circ $ and $ \bullet $ denote the estimates
in the bases (a) and (b), respectively,
and $ \epsilon^h_i = ( m_{q_i} / m_D ) \epsilon^f_i $
as Eq. (\ref{eqn:(a)-(b)}) is adopted in the basis (a)
though $ \epsilon^h_i = 0 $ formally.
This plot of $ \Gamma_D (Z) $ spreads
according to the variation of $ m_D \sim 100 {\rm GeV} - 1 {\rm TeV} $
due to the fact that $ \Gamma_D (Z) $ is almost proportional to $ m_D^3 $.
%%%%%
%%%%%

%%%%%
\begin{figure}
\begin{center}
\scalebox{.5}{\includegraphics*[0.5cm,1cm][15cm,14cm]{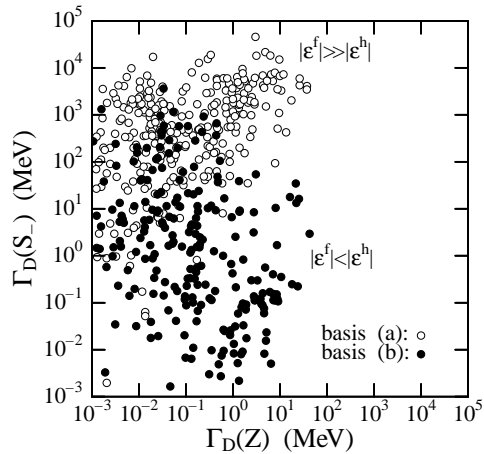}}
\caption{$ \Gamma_D (Z) $ and $ \Gamma_D ( S_- ) $
are compared in the bases (a) ($ \circ $) and (b) ($ \bullet $).}
\label{GZGS}
\end{center}
\end{figure}
%%%%%
%%%%%
\begin{figure}
\begin{center}
\scalebox{.5}{\includegraphics*[0.5cm,1cm][15cm,14cm]{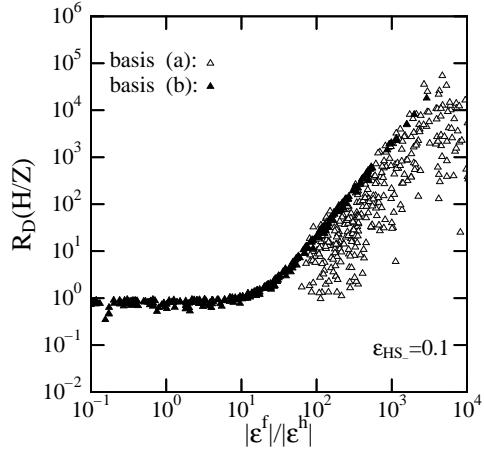}}
\caption{
$ R_D (H/Z) \equiv \Gamma_D (H) / \Gamma_D (Z) $
versus $ | \epsilon^f | / | \epsilon^h | $
is shown for the bases (a) ($ \vartriangle $) and (b) ($ \blacktriangle $),
where $ \epsilon_{HS_-} = 0.1 $ is taken
typically for the $ H $-$ S_- $ mixing.}
\label{efehGHZ}
\end{center}
\end{figure}
%%%%%
The decay widths $ \Gamma_D (Z) $ and $ \Gamma_D ( S_- ) $
for the significant modes are compared in Fig. \ref{GZGS}.
According to Eqs. (\ref{eqn:GDZ}) and (\ref{eqn:GDSm}),
by measuring these decay widths
we can estimate the magnitudes of $ d $-$ D $ mixings,
$ | \epsilon^h | $ from the $ h_d $ coupling
and $ | \epsilon^f | $ from the $ f_D $ and $ f_D^\prime $ couplings.
Specifically, $ \Gamma_D ( S_- ) \gg \Gamma_D (Z) $
for $ | \epsilon^f | \gg | \epsilon^h | $
as in the basis (a) ($ \circ $),
while $ \Gamma_D ( S_- ) \lesssim \Gamma_D (Z) $
for $ | \epsilon^f | \lesssim | \epsilon^h | $
as in the basis (b) ($ \bullet $).
The decay width $ \Gamma_D (H) $ with the standard Higgs $ H $
is also relevant to measure the relative significance
of $ | \epsilon^h | $ versus $ | \epsilon^f | $
according to Eq. (\ref{eqn:GDSmH})
with the sizable $ H $-$ S_- $ mixing.
This is useful even if the decays $ D \rightarrow d_i S_- $
are kinematically forbidden for $ m_{S-} > m_D + m_{d_i} $.
A plot of $ R_D (H/Z) $ versus $ | \epsilon^f | / | \epsilon^h | $
is shown in Fig. \ref{efehGHZ} for the bases (a) ($ \vartriangle $)
and (b) ($ \blacktriangle $),
where $ \epsilon_{HS_-} = 0.1 $ is taken typically
for the $ H $-$ S_- $ mixing.
In the region of $ | \epsilon^f | / | \epsilon^h | \gg 1 $,
the contribution of the singlet Higgs coupling $ \Lambda_{\cal D}^{S_-} $
dominates as $ \Gamma_D (H) \sim \epsilon_{H S^-}^2 \Gamma_D ( S_- )
\gg \Gamma_D (Z) $.
On the other hand,
in the region of $ | \epsilon^f | / | \epsilon^h | \lesssim 1 $
we have $ \Gamma_D (H) \sim \Gamma_D (Z) $ as usually expected.

In these plots of Figs. \ref{mDGWHZ}, \ref{ehGZ}, \ref{GZGS}
and \ref{efehGHZ}, the regions for the bases (a) and (b)
overlap as expected, but they are not identical.
This is because the actual parameter ranges
are somewhat different for the bases (a) and (b);
although the parameter ranges have been taken
apparently in the same way for these bases
in the numerical calculations,
except that $ h_q = {\bf 0} $ in the basis (a),
they are not mapped identically to each other
by the unitary transformation between the bases (a) and (b).
Specifically, in the basis (a) we have a significant constraint
$ | \epsilon^h | / | \epsilon^f | \lesssim m_b / m_D \sim 0.01 $
from the relation $ \epsilon^h_i \sim ( m_{d_i} / m_D ) \epsilon^f_i $,
implying $ | \epsilon^h | \lesssim 0.01 $
as long as $ | \epsilon^f | \lesssim 1 $.
This is explicitly seen in Figs. \ref{ehGZ} and \ref{efehGHZ}.
We also note that the plot for the basis (a) in Fig. \ref{efehGHZ}
spreads significantly.
This is in some sense an artifact
due to the definition of the $ d $-$ D $ mixing parameters
$ \epsilon^f_i $ in terms of $ f_D^+ \equiv f_D + f^\prime_D $
for the basis (a).
The singlet $ S_- $ coupling $ \Lambda_{\cal D}^{S_-} $
is rather given by $ f_D^- \equiv i ( f_D - f^\prime_D ) $.
The spread in the plot for the basis (a)
really reflects the partial cancellation between $ f_D $ and $ f^\prime_D $
for the $ \Lambda_{\cal D}^{S_-} $ coupling.
On the other hand, for the basis (b)
the parameters $ \epsilon^f_i $ are defined suitably
with $ f_D^- = 2i f_D $ ($ f_D = - f^\prime_D $ for $ f_D^+ = {\bf 0} $).
Hence, the plot for the basis (b) almost lies
on a curve up to the small fluctuation due to the kinematic factor,
which gives the boundary of the plot for the basis (a).
This boundary really corresponds to the extreme case
$ f_D \approx - f^\prime_D $ for $ | f_D^- | \approx 2 | f_D | $
in the basis (a).

As seen so far, the singlet $ D $ quark decays present us important insights
on the $ d $-$ D $ mixing effects for the flavor-changing processes.
Especially, if it is observed that $ \Gamma_D ( S_- ) \gg \Gamma_D (Z) $,
we find that the singlet Higgs scalar interactions
dominate over the $ Z $-boson interactions.
For example, suppose that the current experimental bound
for the $ B_d $-$ {\bar B_d} $ mixing \cite{UTfit} is almost saturated
with $ ( \epsilon^f_1 \epsilon^f_3 )^{\frac{1}{2}} \sim 0.2 $
for $ m_{S_-} \sim 300 {\rm GeV} $ in the basis (a)
as shown in Eq. (\ref{eqn:efLmSma}),
which implies $ | \epsilon^f | \gtrsim {\sqrt 2} \times 0.2 
\gg | \epsilon^h | $.
Then, we expect $ \Gamma_D ( S_- ) \sim 1 {\rm GeV} - 10 {\rm GeV}
\gg \Gamma_D (Z) $ for $ m_D \sim 500 {\rm GeV} - 1 {\rm TeV} $,
as seen in Eq. (\ref{eqn:GDSm}) and Fig. \ref{GZGS}.
Contrarily, if $ \Gamma_D ( S_- ) \lesssim 1 {\rm MeV} $
for $ m_D \gtrsim 500 {\rm GeV} $, which implies
$ | \epsilon^f | \lesssim 0.01 $,
the scalar FCNC's do not provide
significant contributions to the $ \Delta F = 2 $ meson mixings.
As for the $ D $ decays with the $ Z $ boson,
if there is a significant left-handed $ d $-$ D $ mixing
as $ | \epsilon^h | \sim 0.03 $ in the basis (b),
we expect $ \Gamma_D (Z) \sim 10 {\rm MeV} - 100 {\rm MeV} $
for $ m_D \sim 500 {\rm GeV} - 1 {\rm TeV} $ in Eq. (\ref{eqn:GDZ}).
In this case, the bound for the $ B_d $-$ {\bar B_d} $ mixing
may be saturated by the $ Z $-mediated FCNC
with $ ( \epsilon^h_1 \epsilon^h_3 )^{\frac{1}{2}} \sim 0.03 $,
as shown in Eq. (\ref{eqn:ehDZb}).
On the other hand, if $ \Gamma_D (Z) \lesssim 0.01 {\rm MeV} $
for $ m_D \gtrsim 500 {\rm GeV} $,
which implies $ | \epsilon^h | \lesssim 0.001 $ (see Fig. \ref{ehGZ}),
the effects of the $ Z $-mediated FCNC's are negligible
in the $ \Delta F = 2 $ meson mixings.

%%%%%
\subsection{Higgs particle decays}
\label{subsec:s-decays}
%%%%%

We next survey the decays of the Higgs particles
$ H $, $ S_+ $ and $ S_- $,
or more precisely the mass eigenstates $ \phi_1 $, $ \phi_2 $, $ \phi_3 $
with the mixing matrix $ O_\phi $.

The standard Higgs $ H $ is probably lighter than the singlet quarks $ Q $
($ m_Q > m_H \approx 120 {\rm GeV} $)
so that its decays involving the singlet quarks
are forbidden kinematically.
It should also be noted
that the $ q $-$ Q $ mixing effect on the $ H $ coupling
with the ordinary quarks appears merely at the second order
related to the modification of the $ Z $-boson coupling,
as seen in Eq. (\ref{eqn:lmH-DZQ}).
Hence, the Higgs particle $ H $ will decay essentially
in the same way as the standard model
unless the $ H $-$ S_- $ mixing is so large
as to provide significant effects.

The singlet Higgs particles $ S_\pm $ will be produced significantly
by gluon fusion via a loop of singlet quark $ Q $ coupled to $ S_\pm $
with the strength $ \sim | \lambda_Q^\pm | \sim m_Q / v_S \sim 1 $.
The production rates of $ S_\pm $
will be comparable to that of the standard Higgs $ H $
unless $ S_\pm $ are substantially heavier than $ H $.
If $ m_{S_\pm} < m_Q $, the singlet quark decays
$ Q \rightarrow q_i S_\pm $ also produce $ S_\pm $, as discussed so far.
It should be remarked
that some indirect indication for the presence of $ S_- $
may be obtained via the $ H $-$ S_- $ mixing,
specifically in the case of $ \Gamma_Q (H) \gg \Gamma_Q (Z) $
for the singlet quark decays $ Q \rightarrow q H $.

In the case of $ m_{S_\pm} < m_Q $,
the singlet Higgs particles $ S_\pm $ decay predominantly
into the ordinary quarks through the scalar interactions
in Eqs. (\ref{eqn:LmSp}), (\ref{eqn:LmSma}) and (\ref{eqn:LmSmb})
at the second order of the $ q $-$ Q $ mixing:
%%%%%
\begin{eqnarray}
S_\pm \rightarrow q_i {\bar q}_j .
\end{eqnarray}
%%%%%
The decay widths are estimated particularly for $ S_- $
in comparison with that of the standard Higgs $ H $ as
%%%%%
\begin{eqnarray}
\frac{\Gamma ( S_- \rightarrow q_i {\bar q}_j )}
{\Gamma ( H \rightarrow b {\bar b} )}
& \sim & \frac{m_{S_-} ( | \Lambda_{\cal Q}^{S_-} [q^c q]_{ij} |^2
+ | \Lambda_{\cal Q}^{S_-} [q^c q]_{ji} |^2 )}
{m_H | \Lambda_{\cal Q}^H [q^c q]_{33} |^2}
\nonumber \\
& \sim & [ ( m_{S_-} / m_H ) ( m_Q / v_S )^2 / ( m_b / v )^2 ]
\nonumber \\
&{}& \times [ ( \epsilon^f_i )^2 ( \epsilon^h_j )^2
+ ( \epsilon^f_j )^2 ( \epsilon^h_i )^2 ] ,
\label{eqn:Smqq}
\end{eqnarray}
%%%%%
where $ \epsilon^h_j \rightarrow ( m_{q_j} / m_Q ) \epsilon^f_j $
in the basis (a).
We estimate, for instance, $ \Gamma ( S_- \rightarrow b {\bar b} ) /
\Gamma ( H \rightarrow b {\bar b} )
\sim [ 10 ( \epsilon^f_3 \epsilon^h_3 )^{\frac{1}{2}} ]^4 $
for $  m_{S_-} / m_H \sim 3 $, $ m_D / v_S \sim 1 $
and $ m_b / v \simeq 1/60 $,
which may amount to $ O (1) $ for the large $ b $-$ D $ mixing
as $ \epsilon^f_3 \sim 1 $
and $ \epsilon^h_3 \sim ( m_b / m_D ) \epsilon^f_3 \sim 0.01 $.
The flavor-changing decays such as $ S_- \rightarrow b {\bar s} $
as well as the flavor-diagonal ones may have significant fractions.
This is distinct from the standard Higgs $ H $,
presenting a promising signature of the singlet Higgs $ S_- $.
In fact, we estimate
%%%%%
\begin{eqnarray}
\frac{\Gamma ( S_- \rightarrow b {\bar d}_j )}
{\Gamma ( S_- \rightarrow b {\bar b} )}
& \sim & ( \epsilon^f_j / \epsilon^f_3 )^2
+ ( \epsilon^h_j / \epsilon^h_3 )^2 ,
\label{eqn:BrSmqq}
\end{eqnarray}
%%%%%
depending on the flavor structure of the $ d $-$ D $ mixing.
If the singlet $ U $ quarks are present with a large $ t $-$ U $ mixing,
the decays $ S_- \rightarrow t {\bar t} , t {\bar u}_i , u_i {\bar t} $
involving the top quark may be observed with significant fractions.

In this way, the decays of the singlet Higgs $ S_- $
into the ordinary quarks are determined
in terms of the $ q $-$ Q $ mixing parameters
with close connection to the flavor-changing processes
such as the $ \Delta F = 2 $ meson mixings.
As for the the singlet Higgs $ S_+ $, its coupling is given
in Eq. (\ref{eqn:LmSp}) by the $ Z $-boson coupling
at the second order of $ q $-$ Q $ mixing
with further suppression by the ordinary quark mass.
These arguments on the $ S_\pm $ couplings with the ordinary quarks
generally suggest that
%%%%%
\begin{eqnarray}
\Gamma_H \gtrsim \Gamma_{S-} \gg \Gamma_{S_+}
( m_{S_\pm} < m_Q , O_\phi \approx {\bf 1} ) .
\end{eqnarray}
%%%%%
for the decay rates of the Higgs particles
if the Higgs mixing is negligibly small.
It is, however, possible that the large Higgs mixing,
in cooperation with the $ q $-$ Q $ mixing,
affects significantly the decays of $ H $ and $ S_\pm $.
(See also Ref. \cite{exHiggs}
for investigations of extended Higgs models at the LHC.)
Therefore, the observations of the Higgs particle decays
present important information on the Higgs mixing
and $ q $-$ Q $ mixing.

In the case of $ m_{S_\pm} > m_Q $,
the singlet quark decays $ Q \rightarrow q S_\pm $
are forbidden kinematically.
Even in such a case the singlet Higgs $ S_\pm $
will be produced significantly
by the gluon fusion via the singlet quark loop.
Then, they decay predominantly involving the singlet quarks as
%%%%%
\begin{eqnarray}
S_\pm \rightarrow Q {\bar q} , {\bar Q} q , Q {\bar Q} .
\end{eqnarray}
%%%%%
The decay widths are estimated in terms of the scalar couplings
$ \Lambda_{\cal Q}^{S_\pm} $ in Eq. (\ref{eqn:LmS}).
In particular, if $ m_{S_\pm} > 2 m_Q $
we have
%%%%%
\begin{eqnarray}
\Gamma ( S_\pm \rightarrow Q {\bar Q} )
\sim \frac{( m_Q / v_S )^2 m_{S_\pm}}{16 \pi} \gtrsim 10 {\rm GeV}
\gg \Gamma_H
\end{eqnarray}
%%%%%
with $ {\rm Br} ( S_\pm \rightarrow Q {\bar Q} ) \approx 1 $
for $ m_{S_\pm} \gtrsim 500 {\rm GeV} $
and $ | \lambda_Q^{\pm} | \sim m_Q / v_S \sim 1 $.

%%%%%%%%%%
\section{Summary}
\label{sec:summary}
%%%%%%%%%%

The singlet quarks in cooperation with the single Higgs field
may provide various interesting effects in particle physics and cosmology
through the mixing with the ordinary quarks ($ q $-$ Q $ mixing).
It is hence worth considering their phenomenological implications
toward the discovery of them at the LHC.
In this study we have investigated the flavor-changing interactions
in the model with singlet quarks and singlet Higgs,
which are induced by the $ q $-$ Q $ mixing.
While the gauge interactions have been investigated extensively
in the literature,
we have rather noted here that the scalar interactions mediated
by the singlet Higg may provide significant effects in some cases.
This possibility has not been paid so much attention before
in the models with singlet quarks.
We have considered the effects of the gauge and scalar interactions
in the $ \Delta F = 2 $ mixings of the neutral mesons
to show the currently allowed range of the $ q $-$ Q $ mixing.
Then, we have investigated the decays of the singlet quarks
and Higgs particles as the new physics
around the electroweak scale to the TeV scale,
which is accessible to the LHC.
Especially, the right-handed $ q $-$ Q $ mixing may be tolerably large
without contradicting the current bounds on the flavor-changing processes,
since it is not involved directly in the electroweak gauge interactions.
If this is the case, the scalar coupling by the singlet Higgs,
and possibly through the Higgs mixing, provides distinct signatures
for the decays of the singlet quarks and Higgs particles,
which should be compared with the conventionally expected ones
via the gauge and standard Higgs couplings.
We expect that observations of the singlet quarks and Higgs particles
will present us important insights
on the $ q $-$ Q $ mixing and Higgs mixing.

%%%%%%%%%%
\acknowledgments
%%%%%%%%%%

We would like to thank M. Senami for valuable discussions.

%%%%%%%%%%
\appendix
%%%%%%%%%%

%%%%%%%%%%
\section{Relations among the gauge and scalar couplings}
\label{sec:couplings}
%%%%%%%%%%

We here derive the suitable relations
among the gauge and scalar couplings.

The full mixing matrix for the left-handed $ W $-boson coupling
is given by
%%%%%
\begin{eqnarray}
{\cal V} &=& {\cal V}_{{\cal U}_{\rm L}}^\dagger
\left( \begin{array}{cc} V_0 & {\bf 0} \\ {\bf 0} & {\bf 0}
\end{array} \right) {\cal V}_{{\cal D}_{\rm L}}
\nonumber \\
&=& \left( \begin{array}{cc} V_{u_{\rm L}}^\dagger V_0 V_{d_{\rm L}}
& V_{u_{\rm L}}^\dagger V_0 \epsilon_{d_{\rm L}} \\
\epsilon_{u_{\rm L}}^\dagger V_0 V_{d_{\rm L}}
& \epsilon_{u_{\rm L}}^\dagger V_0 \epsilon_{d_{\rm L}}
\end{array} \right) .
\end{eqnarray}
%%%%%
Then, we obtain from the off-diagonal blocks
%%%%%
\begin{eqnarray}
{\cal V} [u^\dagger D]
&=& V_{u_{\rm L}}^\dagger V_0 \epsilon_{d_{\rm L}}
\simeq V V_{d_{\rm L}}^\dagger \epsilon_{d_{\rm L}} ,
\label{eqn:VuD-A}
\\
{\cal V} [U^\dagger d]
&=& \epsilon_{u_{\rm L}}^\dagger V_0 V_{d_{\rm L}}
\simeq \epsilon_{u_{\rm L}}^\dagger V_{u_{\rm L}} V ,
\label{eqn:VdU-A}
\end{eqnarray}
%%%%%
where the approximate unitarity
$ V_{q_{\rm L}} V_{q_{\rm L}}^\dagger \simeq {\bf 1} $
is considered up to the second order of the small $ q $-$ Q $ mixing.
By applying Eq. (\ref{eqn:ZqQ}) for the $ Z $-boson coupling
to Eqs. (\ref{eqn:VuD-A}) and (\ref{eqn:VdU-A}),
we obtain Eqs. (\ref{eqn:VuD}) and (\ref{eqn:VdU}).

The left-handed $ Z $-boson coupling is given originally
in the electroweak basis $ ( q_0 , Q_0 ) $ as
%%%%%
\begin{eqnarray}
{\cal Z}_{\cal Q}^{(0)}
= \left( \begin{array}{cc} {\bf 1} & {\bf 0} \\ {\bf 0} & {\bf 0}
\end{array} \right)
- a \left( \begin{array}{cc} {\bf 1} & {\bf 0} \\ {\bf 0} & {\bf 1}
\end{array} \right) ,
\label{eqn:ZQ0}
\end{eqnarray}
%%%%%
where $ a = \sin^2 \theta_W Q_{\rm em} ({\cal Q}) / I_3 (q_0) $,
and the division by $ I_3 (q_0) = \pm 1/2 $ is for convenience of notation.
It is transformed in the basis of mass eigenstates $ ( q , Q ) $ as
%%%%%
\begin{eqnarray}
{\cal Z}_{\cal Q}
&=& {\cal V}_{{\cal Q}_{\rm L}}^\dagger
{\cal Z}_{\cal Q}^{(0)} {\cal V}_{{\cal Q}_{\rm L}}
\nonumber \\
&=& {\cal V}_{{\cal Q}_{\rm L}}^\dagger
\left( \begin{array}{cc} {\bf 1} & {\bf 0} \\ {\bf 0} & {\bf 0}
\end{array} \right) {\cal V}_{{\cal Q}_{\rm L}}
- a \left( \begin{array}{cc} {\bf 1} & {\bf 0} \\ {\bf 0} & {\bf 1}
\end{array} \right) .
\label{eqn:ZQ}
\end{eqnarray}
%%%%%
The modification of the $ Z $-boson coupling
due to the $ q $-$ Q $ mixing is calculated
from Eqs. (\ref{eqn:ZQ0}) and (\ref{eqn:ZQ}) as
%%%%%
\begin{eqnarray}
\Delta {\cal Z}_{\cal Q}
&=& {\cal Z}_{\cal Q} - {\cal Z}_{\cal Q}^{(0)}
\nonumber \\
&=& {\cal V}_{{\cal Q}_{\rm L}}^\dagger
\left( \begin{array}{cc} {\bf 1} & {\bf 0} \\ {\bf 0} & {\bf 0}
\end{array} \right) {\cal V}_{{\cal Q}_{\rm L}}
- \left( \begin{array}{cc} {\bf 1} & {\bf 0} \\ {\bf 0} & {\bf 0}
\end{array} \right)
\nonumber \\
&=& \left( \begin{array}{cc}
- \epsilon_{q_{\rm L}}^\prime \epsilon_{q_{\rm L}}^{\prime \dagger}
& V_{q_{\rm L}}^\dagger \epsilon_{q_{\rm L}} \\
\epsilon_{q_{\rm L}}^\dagger V_{q_{\rm L}}
& \epsilon_{q_{\rm L}}^\dagger \epsilon_{q_{\rm L}} \end{array} \right) .
\label{eqn:DZQ-A}
\end{eqnarray}
%%%%%
Here, we have considered the relation
$ V_{q_{\rm L}}^\dagger V_{q_{\rm L}} - {\bf 1}
= - \epsilon_{q_{\rm L}}^\prime \epsilon_{q_{\rm L}}^{\prime \dagger} $
from the unitarity of $ {\cal V}_{{\cal Q}_{\rm L}} $
to obtain Eq. (\ref{eqn:DZQ})
for the upper diagonal block $ \Delta {\cal Z}_{\cal Q} [ q^\dagger q ] $
in $ \Delta {\cal Z}_{\cal Q} $.
We also obtain the $ Z $ boson $ q $-$ Q $ couplings in Eq. (\ref{eqn:ZqQ})
from the off-diagonal blocks in Eqs. (\ref{eqn:ZQ}) and (\ref{eqn:DZQ-A})
as
%%%%%
\begin{eqnarray}
{\cal Z}_{\cal Q} [q^\dagger Q]
= \Delta {\cal Z}_{\cal Q} [q^\dagger Q]
= V_{q_{\rm L}}^\dagger \epsilon_{q_{\rm L}}
= ( {\cal Z}_{\cal Q} [Q^\dagger q] )^\dagger .
\label{eqn:ZqQ-A}
\end{eqnarray}
%%%%%

We note the relation from Eq. (\ref{eqn:DZQ-A}) as
%%%%%
\begin{eqnarray}
{\cal V}_{{\cal Q}_{\rm L}}^\dagger
\left( \begin{array}{cc} {\bf 1} & {\bf 0} \\ {\bf 0} & {\bf 0}
\end{array} \right) {\cal V}_{{\cal Q}_{\rm L}}
= \Delta {\cal Z}_{\cal Q}
+ \left( \begin{array}{cc} {\bf 1} & {\bf 0} \\ {\bf 0} & {\bf 0}
\end{array} \right) .
\label{eqn:VQd10VQ}
\end{eqnarray}
%%%%%
By taking the difference between this relation
and that for the unit matrix
%%%%%
\begin{eqnarray}
{\cal V}_{{\cal Q}_{\rm L}}^\dagger
\left( \begin{array}{cc} {\bf 1} & {\bf 0} \\ {\bf 0} & {\bf 1}
\end{array} \right) {\cal V}_{{\cal Q}_{\rm L}}
= \left( \begin{array}{cc} {\bf 1} & {\bf 0} \\ {\bf 0} & {\bf 1}
\end{array} \right) ,
\end{eqnarray}
%%%%%
we obtain another relation
%%%%%
\begin{eqnarray}
{\cal V}_{{\cal Q}_{\rm L}}^\dagger
\left( \begin{array}{cc} {\bf 0} & {\bf 0} \\ {\bf 0} & {\bf 1}
\end{array} \right) {\cal V}_{{\cal Q}_{\rm L}}
= - \Delta {\cal Z}_{\cal Q}
+ \left( \begin{array}{cc} {\bf 0} & {\bf 0} \\ {\bf 0} & {\bf 1}
\end{array} \right) .
\label{eqn:VQd01VQ}
\end{eqnarray}
%%%%%
By using Eq. (\ref{eqn:VQd10VQ}), we calculate
%%%%%
\begin{eqnarray}
{\cal V}_{{\cal Q}_{\rm R}}^\dagger
{\cal M}_{\cal Q}
\left( \begin{array}{cc} {\bf 1} & {\bf 0} \\ {\bf 0} & {\bf 0}
\end{array} \right) {\cal V}_{{\cal Q}_{\rm L}}
&=& {\bar{\cal M}}_{\cal Q}
{\cal V}_{{\cal Q}_{\rm L}}^\dagger
\left( \begin{array}{cc} {\bf 1} & {\bf 0} \\ {\bf 0} & {\bf 0}
\end{array} \right) {\cal V}_{{\cal Q}_{\rm L}}
\nonumber \\
&=& {\bar{\cal M}}_{\cal Q} \Delta {\cal Z}_{\cal Q}
+ \left( \begin{array}{cc}
{\bar M}_q & {\bf 0} \\ {\bf 0} & {\bf 0}
\end{array} \right) .
\nonumber \\
&{}&
\label{eqn:VdMQ10VQ-1}
\end{eqnarray}
%%%%%
On the other hand, by considering Eq. (\ref{eqn:subs-MQ1})
for $ M_q $ and $ \Delta_{qQ}^\prime $ in $ {\cal M}_{\cal Q} $
we obtain $ \Lambda_{\cal Q}^H $ in Eq. (\ref{eqn:LmH}) as
%%%%%
\begin{eqnarray}
{\cal V}_{{\cal Q}_{\rm R}}^\dagger
{\cal M}_{\cal Q}
\left( \begin{array}{cc} {\bf 1} & {\bf 0} \\ {\bf 0} & {\bf 0}
\end{array} \right) {\cal V}_{{\cal Q}_{\rm L}}
&=& \frac{v}{\sqrt 2} {\cal V}_{{\cal Q}_{\rm R}}^\dagger
\left( \begin{array}{cc} \lambda_q & {\bf 0} \\ h_q & {\bf 0}
\end{array} \right) {\cal V}_{{\cal Q}_{\rm L}} .
\nonumber \\
&{}&
\label{eqn:VdMQ10VQ-2}
\end{eqnarray}
%%%%%
Comparison of Eqs. (\ref{eqn:VdMQ10VQ-1}) and (\ref{eqn:VdMQ10VQ-2})
establishes the relation between the $ Z $-boson coupling
and the standard Higgs $ H $ coupling,
%%%%%
\begin{eqnarray}
\Lambda_{\cal Q}^H
= ( {\bar{\cal M}}_{\cal Q} / v ) \Delta {\cal Z}_{\cal Q}
+ \left( \begin{array}{cc}
{\bar M}_q / v & {\bf 0} \\ {\bf 0} & {\bf 0}
\end{array} \right) .
\label{eqn:LmHDZ}
\end{eqnarray}
%%%%%
Specifically, Eq. (\ref{eqn:lmH-DZQ})
is obtained from the upper diagonal block,
and Eqs. (\ref{eqn:LHqQ}) and (\ref{eqn:LHQq})
from the off-diagonal blocks
with $ {\cal Z}_{\cal Q} [q^\dagger Q]
= \Delta {\cal Z}_{\cal Q} [q^\dagger Q] $.

Similarly, we obtain the relation
between the $ Z $-boson coupling and the singlet Higgs $ S_+ $ coupling
as follows.
By using Eq. (\ref{eqn:VQd01VQ}), we calculate
%%%%%
\begin{eqnarray}
{\cal V}_{{\cal Q}_{\rm R}}^\dagger
{\cal M}_{\cal Q}
\left( \begin{array}{cc} {\bf 0} & {\bf 0} \\ {\bf 0} & {\bf 1}
\end{array} \right) {\cal V}_{{\cal Q}_{\rm L}}
&=& {\bar{\cal M}}_{\cal Q}
{\cal V}_{{\cal Q}_{\rm L}}^\dagger
\left( \begin{array}{cc} {\bf 0} & {\bf 0} \\ {\bf 0} & {\bf 1}
\end{array} \right) {\cal V}_{{\cal Q}_{\rm L}}
\nonumber \\
&=& - {\bar{\cal M}}_{\cal Q} \Delta {\cal Z}_{\cal Q}
+ \left( \begin{array}{cc}
{\bf 0} & {\bf 0} \\ {\bf 0} & {\bar M}_Q
\end{array} \right) .
\nonumber \\
&{}&
\label{eqn:VdMQ01VQ-1}
\end{eqnarray}
%%%%%
On the other hand, by considering Eq. (\ref{eqn:subs-MQ2})
for $ \Delta_{qQ} $ and $ M_Q $ in $ {\cal M}_{\cal Q} $
we obtain $ \Lambda_{\cal Q}^{S_+} $ in Eq. (\ref{eqn:LmS}) as
%%%%%
\begin{eqnarray}
{\cal V}_{{\cal Q}_{\rm R}}^\dagger
{\cal M}_{\cal Q}
\left( \begin{array}{cc} {\bf 0} & {\bf 0} \\ {\bf 0} & {\bf 1}
\end{array} \right) {\cal V}_{{\cal Q}_{\rm L}}
&=& \frac{v_S}{\sqrt 2} {\cal V}_{{\cal Q}_{\rm R}}^\dagger
\left( \begin{array}{cc} {\bf 0} & f_Q^+ \\ {\bf 0} & \lambda_Q^+
\end{array} \right) {\cal V}_{{\cal Q}_{\rm L}} .
\nonumber \\
&{}&
\label{eqn:VdMQ01VQ-2}
\end{eqnarray}
%%%%%
Comparison of Eqs. (\ref{eqn:VdMQ01VQ-1}) and (\ref{eqn:VdMQ01VQ-2})
leads to the expected relation
%%%%%
\begin{eqnarray}
\Lambda_{\cal Q}^{S_+}
= - ( {\bar{\cal M}}_{\cal Q} / v_S ) \Delta {\cal Z}_{\cal Q}
+ \left( \begin{array}{cc}
{\bf 0} & {\bf 0} \\ {\bf 0} & {\bar M}_Q / v_S
\end{array} \right) .
\label{eqn:LmSDZ}
\end{eqnarray}
%%%%%
Specifically, Eq. (\ref{eqn:LmSp})
is obtained from the upper diagonal block,
and Eqs. (\ref{eqn:LS+qQ}) and (\ref{eqn:LS+Qq})
from the off-diagonal blocks
with $ {\cal Z}_{\cal Q} [q^\dagger Q]
= \Delta {\cal Z}_{\cal Q} [q^\dagger Q] $.

\end{document}